\documentclass{aa}
\usepackage{pdflscape}
\usepackage{graphicx}
\usepackage{txfonts}
\usepackage{textcomp, gensymb}
\usepackage{lscape}
\usepackage[dvipsnames]{xcolor}
\usepackage{placeins}
\usepackage[version=4]{mhchem} 
\usepackage[colorlinks = true,
            linkcolor = blue,
            urlcolor  = blue,
            citecolor = blue,
            anchorcolor = blue, unicode]{hyperref}

\begin{document}

   \title{Understanding JWST water spectra: what can thermochemical models tell us about the (cold) water in protoplanetary disks?}
   \titlerunning{What can thermochemical models tell us about the (cold) water in protoplanetary disks?}

   \author{Marissa Vlasblom\inst{1}
          \and
          Milou Temmink\inst{1}
          \and 
          Andrew D. Sellek\inst{1}
          \and
          Ewine F. van Dishoeck\inst{1,2}
          }

   \institute{Leiden Observatory, Leiden University, 2300 RA Leiden, Netherlands\\ 
              \email{vlasblom@strw.leidenuniv.nl}
    \and
    Max-Planck Institut f\"{u}r Extraterrestrische Physik (MPE), Giessenbachstr. 1, 85748, Garching, Germany 
   }
   \date{Received 4 June 2025; accepted 28 August 2025}
 
  \abstract 
   {Rotational \ce{H2O} spectra as observed with JWST/MIRI trace a wide range of excitation conditions and thereby provide a good probe of the temperature and column density structure of the inner disk. \ce{H2O} emission can also be influenced by dynamical processes in the disk. In particular, dust grains can drift inwards and their icy mantles sublimate once they cross the snowlines, thus enriching the inner regions in, for instance, \ce{H2O} vapor. Recent work has found that this process may leave an imprint in the \ce{H2O} spectrum in the form of excess flux in the cold, low-$E_{\rm up}$ \ce{H2O} lines. }
   {To interpret JWST spectra, LTE slab models are commonly used to determine the temperature, column density, and emitting region that is traced by the observed emission. In this work, we aim to test the accuracy of several common retrieval techniques on full 2D thermochemical disk models, to derive the underlying 2D distribution. Moreover, we investigate the cold \ce{H2O} emission that has been proposed as a signature of drift, to gain further insights into the underlying radial and vertical distribution of \ce{H2O}. }
   {We present two sets of Dust And LInes (DALI) thermochemical models, one where the abundances are set by the chemical network and one where the abundances are parameterized. We run several commonly used retrieval techniques on the generated synthetic spectra and investigate how the retrieved temperature and column density compare to our models. }
   {Single-temperature slab retrievals mainly trace the warm ($\sim$500 K) \ce{H2O} reservoir, whereas a three-component fit is able to better trace the full temperature gradient in the IR emitting region. Retrieved temperatures tend to underestimate the true temperature of the emitting layer due to non-LTE effects such as sub-thermal excitation. The retrieved column density traces close to the mid-IR dust $\tau=1$ surface. The same conclusions are found when performing this analysis for \ce{CO2} emission, and we find that \ce{^13CO2} emission retrieves a lower temperature than \ce{^12CO2} due to it tracing deeper into the disk. Additionally, we find that our fiducial parameterized model predicts a very strong flux in the cold \ce{H2O} lines, but only when the \ce{H2O} abundance in the upper layers is high, whereas the fiducial model with the full chemistry does not. }
   {We find that the strength of the cold \ce{H2O} emission is directly linked to the \ce{H2O} abundance above the snow surface at large radii (>1 au). This implies that sources in which the excess cold \ce{H2O} flux is detected likely have a high \ce{H2O} abundance in this region ($\gtrsim10^{-5}$), higher than what is predicted by the chemical network. This discrepancy is most likely caused by the absence of dust transport processes in our models, further strengthening the theory that this emission may be a signature of radial drift {and vertical mixing}.}

   \keywords{ protoplanetary disks – stars: variables: T Tauri, Herbig Ae/Be – infrared: general – astrochemistry}

   \maketitle
%
\nolinenumbers
\section{Introduction} \label{sec:intro}

Understanding the chemistry of the inner planet-forming region of protoplanetary disks is crucial for understanding the composition and atmospheric properties of exoplanets. Most rocky and super-Earth planets are {proposed} to form in the inner few au of disks, and thus inherit their bulk composition directly from this region \citep[e.g.,][]{dawson2018, oberg2021}. The inner disk regions can be probed with infrared (IR) facilities, and are now being characterized in increasing detail by the \textit{James Webb} Space Telescope (JWST) {\citep[e.g.,][]{kamp2023, henning2024, vandishoeck2023, banzatti2023, romero-mirza2024_sample, gasman2025, arulanantham2025}}. In particular, the characterization of \ce{H2O} emission with JWST is of great interest due to its potential role in creating habitable conditions \citep[see, e.g.,][]{krijt2023_PPVII}. \\
\newline
The rotational \ce{H2O} spectrum as seen with for instance JWST/MIRI covers transitions spanning a wide range of upper level energies $E_{\rm up}$ ($\sim1000-10\,000$ K) and Einstein-A coefficients $A_{\rm ij}$ ($\sim0.01-100$ s$^{-1}$), and hence the lines probe a wide range of excitation conditions \citep[see, e.g.,][]{gasman2023b, banzatti2023, banzatti2025, temmink2024b}. Using the relative excitation of higher and lower $E_{\rm up}$ lines, the \ce{H2O} spectrum can be used as a proxy to retrieve the temperature and \ce{H2O} column density structure of the IR emitting region in the inner disk \citep{romero-mirza2024_sample, temmink2025}. It thus contains very valuable information, both for informing disk modeling efforts and for understanding the conditions under which nascent planets may be forming in the disk.  \\
\newline
\ce{H2O} is produced in the inner disk via high-temperature gas-phase formation at rapid timescales via the \ce{OH + H2 -> H2O + H} reaction, which is efficient at temperatures $\gtrsim$300 K \citep{charnley1997, vandishoeck2013, walsh2015}. In the warm, upper layers of the disk, this reaction is balanced by rapid photodissociation due to the lower dust opacity. In the outer disk (>few au), the temperature becomes low enough ($\lesssim150$ K) for the \ce{H2O} to freeze out onto the dust grains. Above this ice reservoir, only a cold photodesorption layer is present, as the temperature is too low in this region for either gas-phase formation or thermal desorption \citep{woitke2009, hogerheijde2011, vandishoeck2021}. Aside from these chemical processes, the \ce{H2O} vapor in the inner disk is also affected by dynamical processes in the disk, such as the inwards drift of dust. When the icy grains in the outer disk move inwards, their ice mantles can sublimate once the grain crosses a species' snowline. This can enhance the amount of \ce{H2O} vapor in the inner disk, and has been explored using data from the \textit{Spitzer} Space Telescope by \citet{banzatti2020}, who find a correlation between the \ce{H2O} flux and dust disk size. Disks with a smaller size as measured by the millimeter dust continuum, which is thought to be caused by efficient radial drift, are found to have stronger \ce{H2O} fluxes, suggesting a link between dust drift and \ce{H2O} vapor in the inner disk. This correlation has also been theoretically motivated by \citet{kalyaan2021}. The effects of dust drift {and \ce{H2O} delivery to} the inner disk have been further explored {in 1D} by the modeling work of, for example, \citet{mah2023}, \citet{sellek2024}, and \citet{houge2025}{, and the effects of dust evolution on mid-IR lines has also been explored in 2D by, for example, \citet{greenwood2019_dustevol}}. \\
\newline
More recently, this link between dust dynamics and \ce{H2O} emission has been expanded upon with the help of JWST's improved spectral resolution and sensitivity compared with \textit{Spitzer}. \citet{banzatti2023} and \citet{romero-mirza2024_sample} demonstrate that compact dust disks show excess \ce{H2O} flux in the cold, low-$E_{\rm up}$ lines with respect to more extended dust disks and propose that this is a signature of strong drift. The low-$E_{\rm up}$ lines trace temperatures of $\lesssim$200-250 K \citep[see also][]{banzatti2025}, which likely trace close to the \ce{H2O} snowline. A large volume of ice crossing the snowline may thus lead to an increased amount of cold, $\sim$200 K \ce{H2O} vapor which translates to an increase in cold \ce{H2O} flux. \\
\newline
Most work so far has used LTE slab models to characterize the \ce{H2O} emission in JWST/MIRI spectra. These models parameterize the emission using only a temperature $T$, column density $N$ and emitting area $A$, and they can be used in various ways, for example as single-component fits \citep[e.g.,][]{grant2023, tabone2023} and two- or three-component fits \citep[e.g.,][]{temmink2024b, pontoppidan2024}. As these models are used to inform us of the temperature and column densities within the planet-forming zone, understanding the accuracy of these retrievals is key to our understanding of disk chemistry. \\
\newline
2D thermochemical models provide an ideal case to put these retrieval methods to the test. The retrieved temperature and column density provide at most only a 1D view of the disk and thus do not directly probe the underlying 2D distribution of \ce{H2O} vapor in the disk. Thermochemical models do contain the full 2D view of the disk, thereby providing insights that slab models alone cannot give. They can inform us how retrievals may provide indirect information about the underlying 2D abundance structure, and they can inform us of the accuracy of the retrievals by directly comparing the retrieved temperatures ({the excitation temperature} $T_{\rm ex}$) and column densities to the model output (e.g., the kinetic temperature $T_{\rm kin}$). As a further test, we also investigate the retrievals on \ce{CO2} emission (see Sect. \ref{app:CO2}).\\
\newline
The goal of this paper is two-fold: we aim to provide a better understanding of the accuracy of retrieval techniques commonly used on observations, and we also investigate the cold \ce{H2O} lines that have been proposed as signatures of radial drift, to understand what insights thermochemical models can provide on the origin of this emission. The structure of this work is as follows. In Sect. \ref{sec:methods}, we describe our modeling approach using the Dust And LInes (DALI) thermochemical code, as well as the retrieval techniques we use to interpret the synthetic DALI spectra. In Sect. \ref{sec:results} we describe the results of our modeling, retrievals, and our insights into the cold \ce{H2O} emission. We discuss the origin of this cold \ce{H2O} and its potential relation to radial drift further in Sect. \ref{sec:discussion}, and we also provide predictions for observing cold \ce{H2O} lines with future far-IR missions. We summarize our main conclusions in Sect. \ref{sec:conclusions}. 



\section{Methods}\label{sec:methods}
\subsection{Thermochemical models}\label{subsec:methods_DALI}

This work makes use of the 2D thermochemical code Dust And LInes \citep[DALI; see][]{bruderer2012, bruderer2013}. This code consists of three main steps. First, the code calculates the local radiation field and dust temperature at all locations in the disk given an input gas and dust density structure. Second, the gas temperature, chemical abundances of all species in the chemical network, and non-LTE excitation for several specified atoms and molecules (and thus their cooling ability) are calculated self-consistently. Third, a raytracing tool is available to obtain line fluxes, spectra and spectral image cubes. We present two sets of models: in the first set of models, the chemical abundances of all species are set using the chemical network and thermal balance, whereas in the second set the 2D abundances are parameterized, as this allows us to investigate more directly how differences in the abundance structure manifest itself in the spectrum. \\
\begin{figure*}
    \centering
    \includegraphics[width=\linewidth]{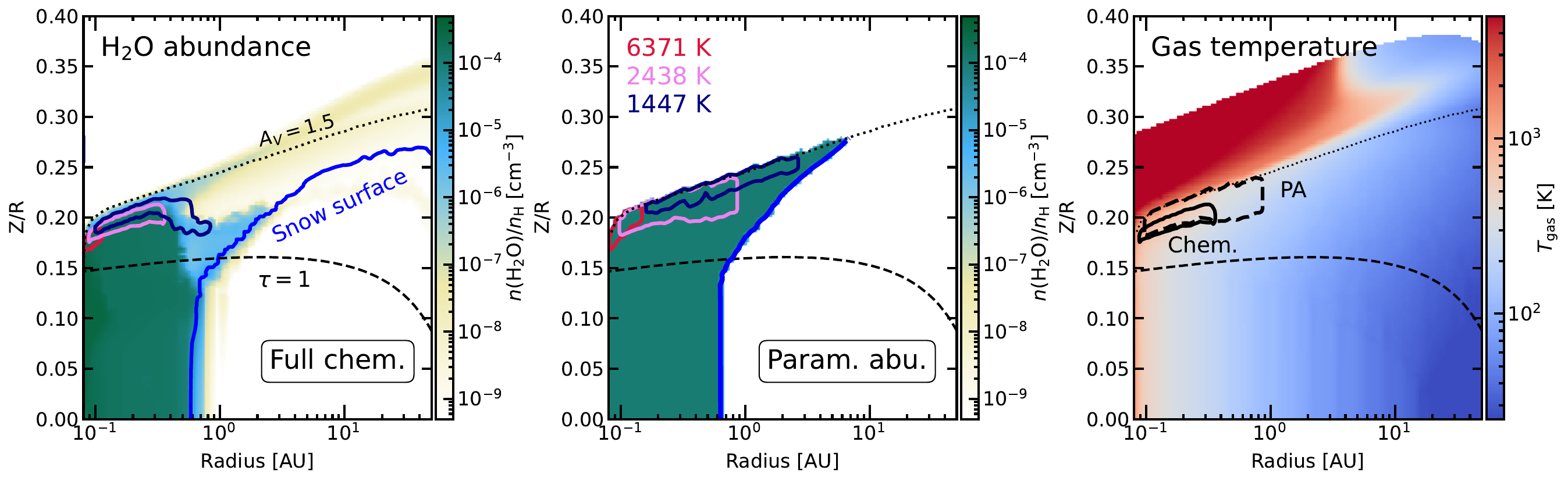}
    \caption{Abundance maps of \ce{H2O} (left \& middle panels) and temperature map (right panel) for the fiducial models with $f_\ell=0.9$. The left panel depicts the model using the chemical network and the middle depicts the model with parameterized abundances (PA). In all panels, the dust $\tau=1$ surface at 15 $\mu$m and the $A_V = 1.5$ surface are indicated with a blue dashed line and a black dotted line, respectively. The red, pink, and blue contours represent the 70\% emitting regions of the \ce{H2O} 17$_{7,10}$ -- 16$_{4,13}$ ($E_{\rm up}$ = 6371 K), 11$_{3,9}$ -- 10$_{0,10}$ ($E_{\rm up}$ = 2438 K), and 8$_{3,6}$ -- 7$_{0,7}$ ($E_{\rm up}$ = 1447 K) lines, respectively, representing hot, warm, and cold \ce{H2O}. The warm \ce{H2O} emitting region for both models is also depicted in the right panel in black. The light blue line in the left and middle panels represents the \ce{H2O} snow surface. }
    \label{fig:abu_fl9}
\end{figure*}
\newline
The input parameters of all models are {summarized in Table \ref{tab:params}. They are} based on those presented in \citet{vlasblom2024}, so we refer the reader to that work and references therein for the full details of the model setup. These models are based on the model by \citet{zhang2021} and \citet{bosman2022a, bosman2022b} for the disk around K5 star AS 209 and therefore use the input stellar spectrum from those works, which has an effective temperature of 4300 K and a bolometric luminosity of 1.4 $L_\odot$. The models contain two dust populations, separated by their grain size. The small dust population has sizes between 5 nm and 1 $\mu$m, and is assumed to be well-coupled to the gas, following the same radial and vertical distribution \citep{miotello2016}. The large dust population has sizes between 5 nm and 1 mm and is assumed to be more settled to the midplane than the small dust, and thus has a reduced scale height. {Both populations follow an MRN distribution \citep{mathis1977}.} The large grains have a mass fraction $f_\ell$, which sets the gas-to-dust ratio in the upper layers of the disk due to the settling of the large grains. The models presented in the main body of this work have $f_\ell=0.9$, which produces a gas-to-dust ratio of $\sim$10$^3$ in the IR emitting region of the disk. This is in line with the predicted elevated gas-to-dust ratio in these layers from \citet{meijerink2009} {and \citet{woitke2018}}. In Sect. \ref{app:fl999}, we discuss the effects of even higher gas-to-dust ratios using models with $f_\ell=0.99$ and 0.999. For our first set of models where we run the chemistry and thermal balance, we use the same chemical network as \citet{bosman2022a, bosman2022b}. This is a network that builds on the standard DALI chemical network which is based on UMIST06 \citep{woodall2007}, by including the effects of \ce{H2O} UV-shielding \citep{bethell2009} and more efficient \ce{H2} formation at high temperatures. We note that the use of a more recent UMIST version, such as UMIST12 or UMIST22, will not affect the results of this work. Isotope chemistry \citep[as implemented in, e.g.,][]{miotello2014, visser2018} is not included in this network.\\
\newline
Aside from the set of models containing the full chemistry, we also present a second set of models where the abundance structure is parameterized using a jump abundance profile, following the methods in \citet{bruderer2015} and \citet{bosman2017}. In these models, the second step of the code (the chemistry and gas thermal balance) is thus skipped, meaning that the gas temperature cannot be calculated self-consistently. Prior works then typically assume it to be equal to the dust temperature. However, in the IR emitting layer, the gas and dust are thermally decoupled and this assumption is not valid \citep[see, e.g.][]{bosman2022a, vlasblom2024}. As such, we use the self-consistently calculated temperature profile from our full chemistry models as the assumed temperature structure for our parameterized models. {This assumption gives a much more realistic gas temperature than setting $T_{\rm gas}=T_{\rm dust}$. However, we do note that our assumed gas temperature for the parameterized models is not entirely self-consistent, as any change in abundance structure could slightly alter the gas temperature as well due to \ce{H2O} being both an effective coolant as well as a source of heating through its photodissociation in the surface layers (see, e.g., \citealt{woitke2018, woitke2024}). } \\
\newline
The parameterized abundance structures are inspired by the chemical network. We parameterize the \ce{H2O} abundance structure using two reservoirs, one inner reservoir with a high fractional abundance of $n$(\ce{H2O})/$n_{\rm H}$ = $10^{-4}$ (where $n_{\rm H}$ = $n$(H) + 2$n$(\ce{H2})) defined as the region where $A_V$ > 1.5 mag and $T_{\rm dust}$ > 150 K. In the rest of the disk, the fractional \ce{H2O} abundance is set to $10^{-10}$. In Sect. \ref{app:CO2}, we show how we parameterize the \ce{CO2} abundance structure. The abundance structure of \ce{H2O} for both the parameterized and full chemistry models are presented in the left column of Fig. \ref{fig:abu_fl9}. \\
\newline
We make use of DALI's ``fast ray-tracer'' \citep[see][Appendix B]{bosman2017} to generate the IR spectra of \ce{H2O} (as well as \ce{^12CO2} and \ce{^13CO2}) between 10 and 28 $\mu$m, assuming a distance of 121 pc and a face-on orientation. Since isotope chemistry is not included in our chemical network, the \ce{^13CO2} spectrum is derived by scaling the \ce{^12CO2} abundance down by a factor of 77, the ISM \ce{^12C/^13C} ratio \citep{wilson1994}. The excitation of the relevant molecules was calculated in non-LTE for both sets of models. The molecular data file for \ce{H2O} is retrieved from the Leiden Atomic and Molecular Database (LAMDA), including levels with energies up to 7200 K \citep{tennyson2001}. The line transitions are obtained from the BT2 list \citep{barber2006} and the collisional rate coefficients come from \citet{faure2008}. For \ce{CO2}, we use the molecular data compiled by \citet{bosman2017}, who retrieve the energy levels, line positions and line strengths from the HITRAN database \citep{rothman2013}, and use collisional rate coefficients based on \citet{allen1980}, \citet{nevdakh2003}, and \citet{jacobs1975}. We compare the non-LTE results with those of an LTE excitation calculation in Sect. \ref{app:LTE}. We convolved the generated spectra to the approximate resolving power of MIRI of $R=3000$, and we do not add synthetic noise. 

\begin{figure*}
    \centering
    \includegraphics[width=\linewidth]{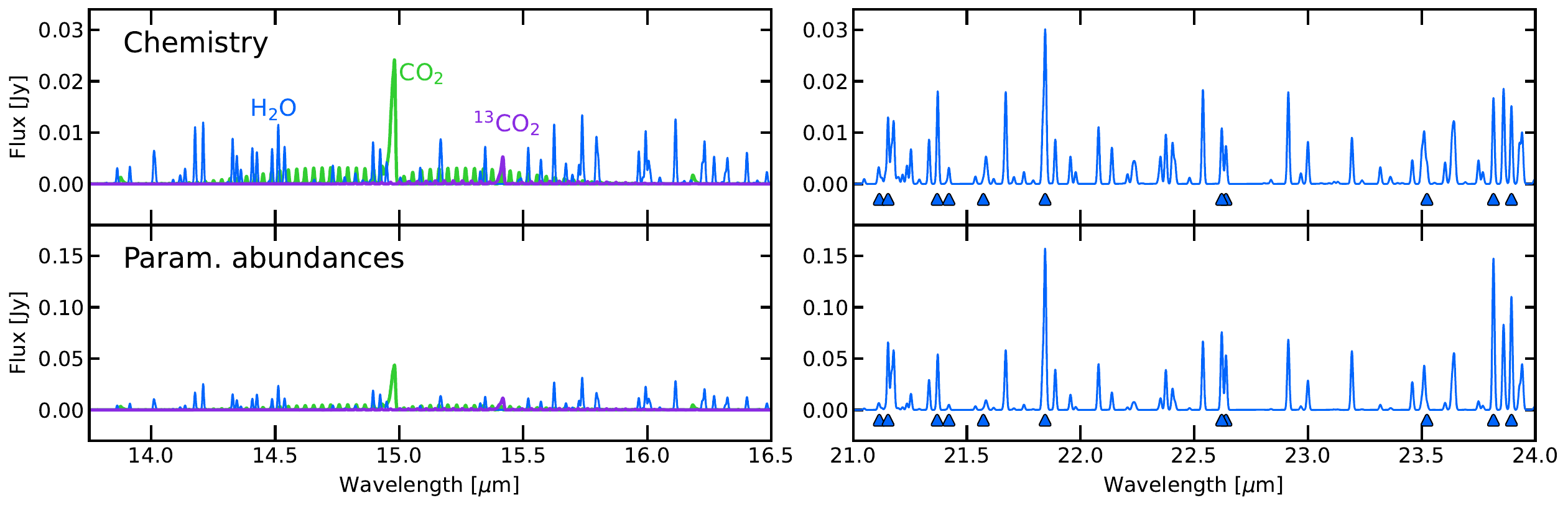}
    \caption{Synthetic \ce{H2O} (blue), \ce{^12CO2} (green) and \ce{^13CO2} (purple) spectra. The top row depicts the model using the chemical network and the bottom row depicts the model with parameterized abundances. In the right panels, blue triangles indicate \ce{H2O} lines with $E_{\rm up} < 2500$ K.}
    \label{fig:spectra_fl9}
\end{figure*}

\subsection{Slab retrievals}\label{subsec:methods_slabs}

Having generated our synthetic spectra, we perform several retrieval analyses using LTE slab models to better understand what those models are tracing in terms of regions and conditions. These slab models contain three free parameters: the gas temperature $T$, the column density $N$, and the emitting area $A$ \citep[see, e.g.,][]{kamp2023}. If one assumes $A = \pi R_{\rm eq}^2$, the emitting area can be converted to an equivalent slab radius {($R_{\rm eq}$)}. This equivalent radius can reflect the true emitting radius of the emission, but it can also originate from an annulus further out in the disk instead. We assume Gaussian line profiles with a full width at half maximum of $\Delta V = 4.7$ km s$^{-1}$, following \citet{salyk2011}. In the main body of this work, we focus on our analysis of the \ce{H2O} emission. We exclude the ro-vibrational band at $\sim$6-8 $\mu$m from our analysis as this band is known to be much more affected by non-LTE effects \citep[see, e.g.,][]{meijerink2009, pontoppidan2024, banzatti2025}. In Sect. \ref{app:CO2}, we present our analysis of the \ce{CO2} emission. \\
\newline
The \ce{H2O} rotational spectrum contains a plethora of lines with a wide range in upper level energy $E_{\rm up}$, thereby tracing a wide range of temperatures and densities \citep[see, e.g.,][]{banzatti2023, banzatti2025}. Thus, fitting the entire rotational spectrum from 10 to 28 $\mu$m with a single temperature is unlikely to provide the best insights. As such, we use two different approaches. \\
\newline
First, we perform single-temperature fits in three different wavelength ranges (10-14, 13.5-17.5, and 21-24 $\mu$m), following works such as \citet{gasman2023b} and \citet{temmink2024b}. As seen in, for instance, \citet{banzatti2023}, the \ce{H2O} lines at the shortest wavelengths ($\sim$10 $\mu$m) generally have the highest $E_{\rm up}$ and should thus trace the highest temperatures in the innermost regions of the disk, with subsequent lines at longer wavelengths having lower $E_{\rm up}$ and thus tracing further out. The best-fit model was obtained from a grid of models using a $\chi^2$ method. In this grid, $T$ was varied linearly in 40 intervals between 100 and 1000 K ($\Delta T = 22.5$ K) and $N$ was varied in log-space in 40 intervals between 10$^{14}$ and 10$^{23}$ cm$^{-2}$ ($\Delta \log N = 0.225$). The spacing of this grid was kept consistent for all fits. For each point in the grid, the best-fit $R_{\rm eq}$ is determined by minimizing the reduced $\chi^2$. We do not account for the effects of mutual shielding of adjacent lines \citep[as done in, e.g.,][]{tabone2023}, as this is also not accounted for when generating our synthetic DALI spectra. \citet{banzatti2025} have demonstrated that, for \ce{H2O}, mutual shielding is important for individual ortho-para line pairs (see their Fig. 5) but does not significantly affect the spectrum as a whole. \citet{temmink2024b} also demonstrate that retrievals are only negligibly affected. \\
\newline
Second, we fit the entire rotational spectrum using three temperature components following scenario I presented in \citet{temmink2024b}. This method combines the flux from three different slab models with decreasing temperature and increasing emitting area, where the best fit is found using the Monte Carlo Markov Chain (MCMC) implementation emcee \citep{foreman_2013_emcee}. The total \ce{H2O} flux is parameterized as 
\begin{align}
\begin{split}
      F_{\rm total} = F_1\pi \left( \frac{R_1}{\mathrm{1\;au}} \right)^2 + &F_2\pi\left[\left( \frac{R_2}{\mathrm{1\;au}} \right)^2 - \left( \frac{R_1}{\mathrm{1\;au}} \right)^2 \right] \\
      + &F_3\pi\left[\left( \frac{R_3}{\mathrm{1\;au}} \right)^2 - \left( \frac{R_2}{\mathrm{1\;au}} \right)^2 \right] 
\end{split}
\end{align}
where $F_i$ and $R_i$ represent the flux and emitting radii of the three slab model components. We do not take into account the effects of spatial shielding of the colder components by the hotter ones (Scenario III in \citealt{temmink2024b}), as this was found to be negligible by \citet{temmink2024b} due to the high optical depth of the hot components. 

\section{Results}\label{sec:results}

\subsection{Abundance maps and synthetic spectra}\label{subsec:res_models}

\begin{figure*}
    \centering
    \includegraphics[width=0.75\linewidth]{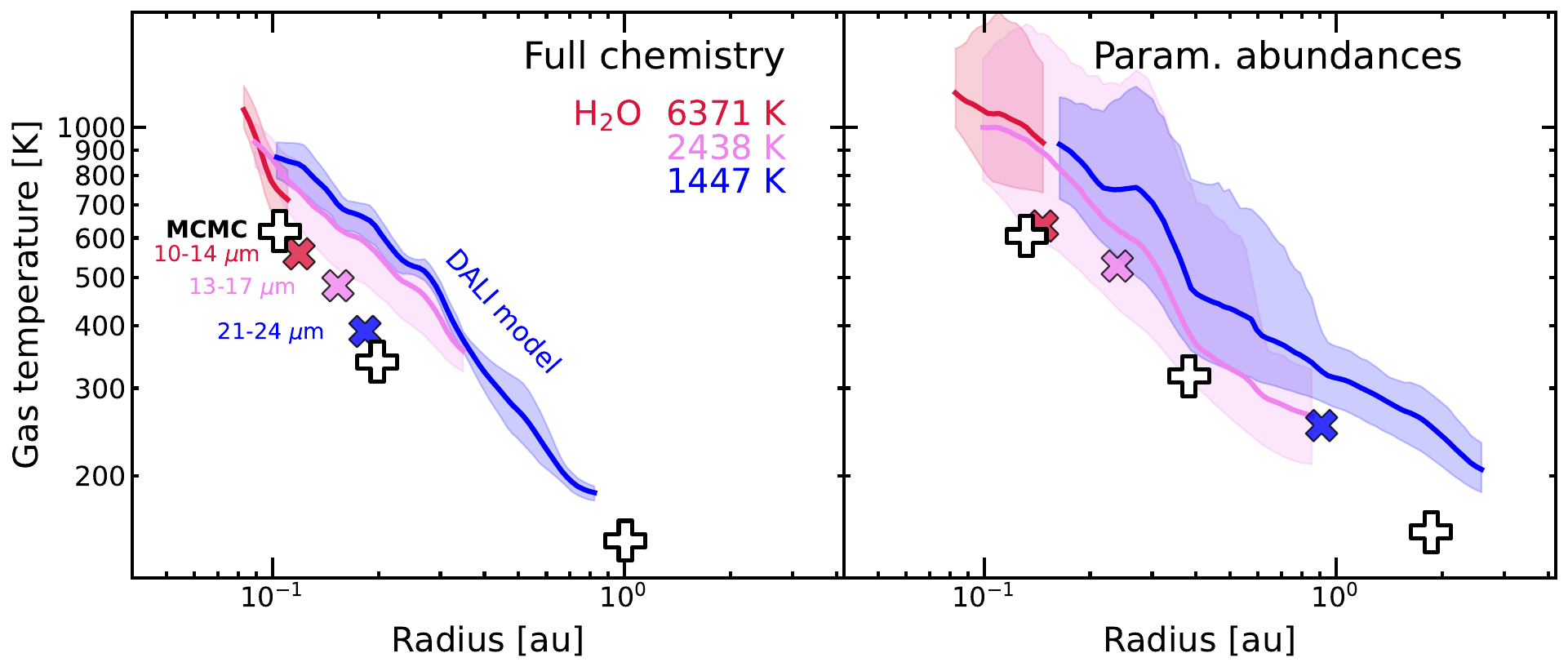}
    \caption{Temperature as a function of radius within the 70\% emitting region of the \ce{H2O} 17$_{7,10}$ -- 16$_{4,13}$ (6371 K), 11$_{3,9}$ -- 10$_{0,10}$ (2438 K), and 8$_{3,6}$ -- 7$_{0,7}$ (1447 K) lines (red, pink, and blue solid lines) for the model using the chemical network (left panel) and the model with parameterized abundances (right panel). The shaded regions represent the minimum and maximum temperature within the emitting region. The red, pink, and blue crosses represent the retrieved $T$ and $R_{\rm eq}$ from the single-temperature slab fits in the 10-14, 13.5-17.5, and 21-24 $\mu$m region. The white plus symbols represent the retrieved $T$ and $R_{\rm eq}$ values from the 3-temperature-component MCMC routine for \ce{H2O}. }
    \label{fig:CBF_Tgas_fl9}
\end{figure*}

We present the \ce{H2O} abundance maps for our full chemistry (left panel) and parameterized abundance models (middle panel), as well as the gas temperature profile (right panel) in Fig. \ref{fig:abu_fl9}. The red, pink and blue contours in the first two panels indicate the line emitting region from which 70\% of the emission originates. These contribution functions (CBFs; see \citealt{bruderer2013} for details on its calculation) are shown for three \ce{H2O} lines with different $E_{\rm up}$, namely the 17$_{7,10}$ -- 16$_{4,13}$ (6371 K; red), 11$_{3,9}$ -- 10$_{0,10}$ (2438 K; pink), and 8$_{3,6}$ -- 7$_{0,7}$ (1447 K; blue) lines (see Table \ref{tab:Lines} for an overview of the properties of all analyzed lines in this work). These lines trace the warm surface layer of the disk, above the $\tau=1$ surface of the dust continuum at 15 $\mu$m. As expected, the lines with higher $E_{\rm up}$ trace hotter regions closer to the star, whereas the lower-$E_{\rm up}$ lines trace slightly further out in the disk, but still within 1 au. \\
\newline
The synthetic spectra for both sets of models are presented in Fig. \ref{fig:spectra_fl9}. In the left panel, the \ce{H2O, ^12CO2}, and \ce{^13CO2} spectra are shown between 13.75 and 16.5 $\mu$m. When comparing the full chemistry and parameterized abundance models, the flux ratios between the three species are similar. The \ce{CO2} emission is stronger than that of \ce{H2O}, which can likely be attributed to the gas-to-dust ratio in the surface layers of these models, which is set to $\sim$$10^3$. \citet{vlasblom2024} demonstrate that higher gas-to-dust ratios ($\sim$$10^4-10^5$) increase the temperature in the surface layers {\citep[see also][]{woitke2018}}, as well as the \ce{H2O/CO2} flux ratio, likely due to the higher temperatures promoting \ce{H2O} production over \ce{CO2}. \\
\begin{figure*}
    \centering
    \includegraphics[width=0.9\linewidth]{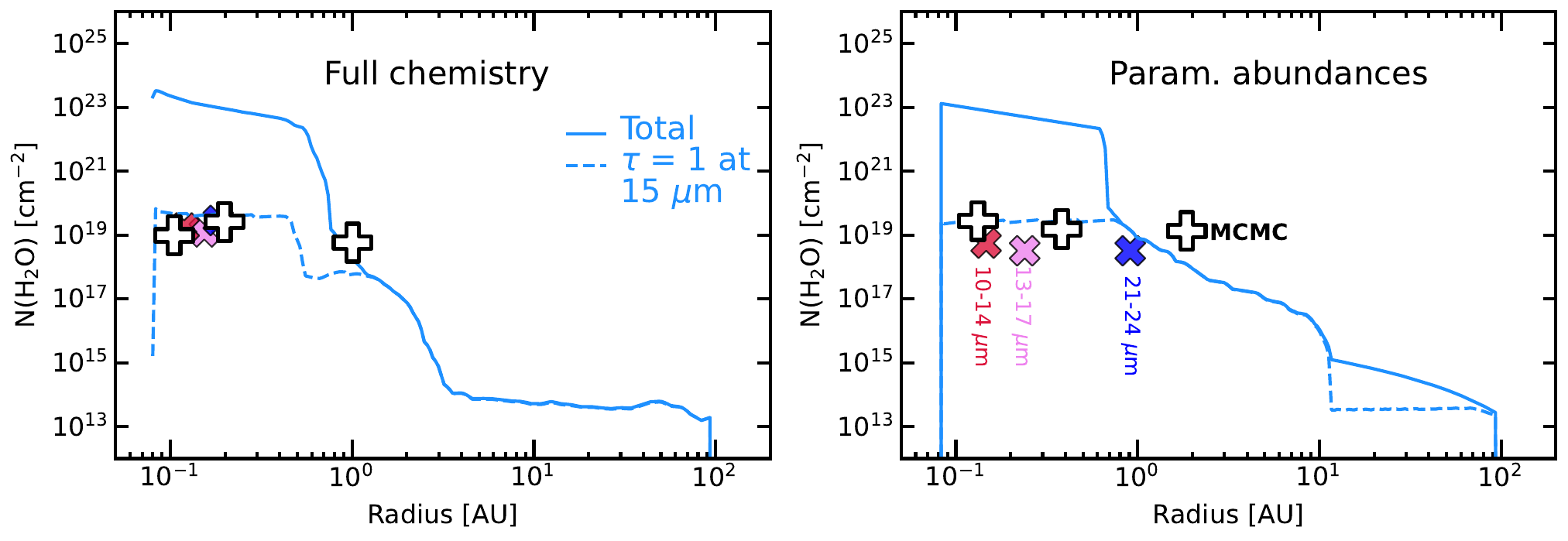}
    \caption{Vertically integrated \ce{H2O} column density as a function of radius. The solid lines show the total model column density and the dashed lines show the model column density integrated up to the dust $\tau=1$ surface at 15 $\mu$m. The red, pink, and blue crosses represent the retrieved $N$ and $R_{\rm eq}$ from the single-temperature slab fits in the 10-14, 13.5-17.5, and 21-24 $\mu$m region. The plus symbols in the left column represent the retrieved $N$ and $R_{\rm eq}$ values from the 3-temperature-component MCMC routine. The left panel depicts the model using the chemical network and the right panel depicts the model with parameterized abundances.}
    \label{fig:N_rad_fl9}
\end{figure*}
\newline
Interestingly, when considering the \ce{H2O} emission between 21 and 24 $\mu$m (right panels of Fig. \ref{fig:spectra_fl9}), the full chemistry and parameterized abundance models are less similar. In the parameterized model, the cold \ce{H2O} emission (lines with $E_{\rm up} < 2500$ K are marked by blue triangles) is strong compared to the surrounding \ce{H2O} lines with higher $E_{\rm up}$. Moreover, the \ce{H2O} lines at $\sim$15 $\mu$m (which are generally dominated by higher $E_{\rm up}$ than those at $\sim$23 $\mu$m) are much weaker than the lines at longer wavelengths, which indicates that the parameterized model is dominated by strong cold \ce{H2O}. In the full chemistry model, the line fluxes at $\sim$15 and $\sim$23 $\mu$m are much more similar, indicative of warmer emission. This difference must be related to the difference in abundance structure between the two models, as all other factors, such as the temperature structure, are controlled for. We discuss this in detail in Sect. \ref{subsec:res_H2O}.

\subsection{Slab retrievals on synthetic spectra}\label{subsec:res_slabs}
\subsubsection{Temperature}\label{subsubsec:res_T}

To compare the results from the slab retrievals to our thermochemical models, we first generate a 1D temperature profile from our 2D temperature structure using the CBFs (see the contours plotted on the abundance maps in Fig. \ref{fig:abu_fl9}). We extract the 1D temperature profile by calculating the CBF-weighted average temperature within the 70\% CBF at each radius, following \citet{kaufer2024}. This is done for all three aforementioned lines. As these lines generally trace similar depths and their radial ranges partially overlap, the three derived temperature profiles closely agree with one another. We compare these temperature profiles to the retrieved values for $T$ and $R_{\rm eq}$ from our slab retrievals in Fig. \ref{fig:CBF_Tgas_fl9}.\\
\newline
As stated in Sect. \ref{subsec:methods_slabs}, the slab models retrieve an emitting area $A$ which can be converted into an equivalent emitting radius assuming $A = \pi R_{\rm eq}^2$, though the emission can also originate from an annulus further out in the disk instead. This can have a significant effect on where the data point will end up on the x-axis of Fig. \ref{fig:CBF_Tgas_fl9}. At a minimum, the inner starting radius of the DALI models, which is set to the dust sublimation radius at 0.08 au, could set the inner radius of such an annulus. However, as can be seen from \ce{CO2} abundance map presented in Fig. \ref{fig:abu_fl9_CO2}, this may not be sufficient in all cases, as the \ce{CO2} emission mainly emits from further out in the disk where it becomes more abundant. As such, we parameterize the retrieved emitting areas from our slab retrievals as an annulus starting at the inner edge of the corresponding 70\% CBF. For the single-temperature \ce{H2O} fits, we match the fit done at the shortest wavelengths to the CBF of the highest $E_{\rm up}$ line, and the subsequent wavelength ranges to the lower $E_{\rm up}$ lines. For the three-temperature-component MCMC, the three components are parameterized as three annuli starting at the inner edge of the CBF of the highest $E_{\rm up}$ line.\\
\newline
As can be seen from Fig. \ref{fig:CBF_Tgas_fl9}, the agreement between the retrieved temperature from the slab fits and the temperature in the emitting layer of the DALI model is generally quite good, for both the full chemistry and parameterized model. It is clear that the three single-temperature fits {(indicated with crosses in Fig. \ref{fig:CBF_Tgas_fl9})} in the 10-14, 13.5-17.5 and 21-24 $\mu$m regions mainly trace the ``warm'', $\sim$500 K component of the emission (though the 21-24 $\mu$m fit in the parameterized model better reflects the cold temperature of this emission, likely due to it being quite strong in this model). While the different wavelength regions do show a slight difference in temperature as expected (colored crosses), they do not capture the full range of temperatures contributing to the emission. Instead, the three-temperature-component MCMC fit does a better job at capturing the gradient in temperature, especially towards the lower temperatures (white crosses). This indicates that a single temperature is not enough to capture the full complexity of the \ce{H2O} spectrum even in more narrow wavelength ranges, due to the wide range of $E_{\rm up}$ and $A_{\rm ij}$ that are present across the rotational spectrum \citep[see, e.g.,][]{meijerink2009, banzatti2025}. Thus, \ce{H2O} retrievals should be done using at least a multi-temperature fit. \\
\newline
Looking closer at the retrieved slab temperatures, while they generally agree well with the temperature in the emitting layer, they are slightly colder. This could be indicative of the lines being slightly sub-thermally excited \citep[see, e.g.,][]{meijerink2009}, as the excitation in the DALI model is calculated in non-LTE whereas the retrieval assumes LTE. We explore the implications of LTE versus non-LTE excitation further in Sect. \ref{app:LTE}, and find that assuming LTE excitation in our DALI model indeed leads to a slightly higher retrieved temperature, though the difference is relatively small ($\sim$100 K).

\subsubsection{Column density}\label{subsubsec:res_N}

Fig. \ref{fig:N_rad_fl9} compares the retrieved column densities from the LTE slab models to the vertically integrated column density. The retrieved values indicate that the emission traces close to the dust $\tau=1$ surface for our fiducial models. In Sect. \ref{app:fl999}, we discuss how these results differ for a set of models with a higher gas-to-dust ratio. In this case, the dust $\tau=1$ surface is located deeper in the disk, whereas the emission does not seem to trace much deeper, given the column densities that are retrieved (see Fig. \ref{fig:N_rad_fl999}). This indicates that, in cases of a high gas-to-dust ratio, the lines can already become optically thick before the dust does, and thus may not trace the dust $\tau=1$ surface in that case.  

\subsection{Investigating the cold \ce{H2O} component}\label{subsec:res_H2O}

Fig. \ref{fig:spectra_fl9} demonstrates that the cold \ce{H2O} emission in the model with parameterized abundances is very strong compared to the warmer emission, whereas that is not the case for the model with the full chemistry. We explore here why this could be the case. 

\subsubsection{Influence of the abundance structure}
\begin{figure}
    \centering
    \includegraphics[width=\linewidth]{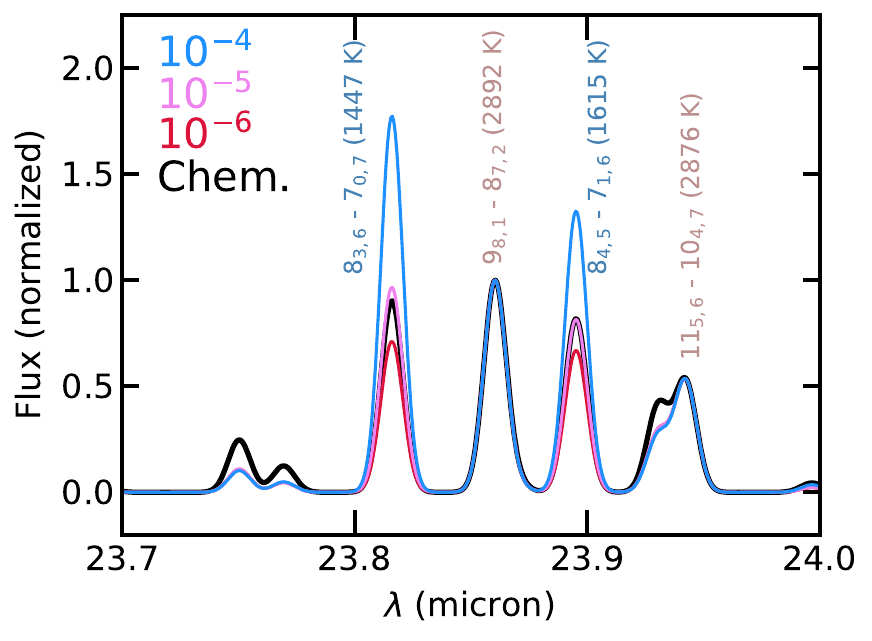}
    \caption{Synthetic \ce{H2O} spectra of the fiducial model using the chemical network (black) and three models with parameterized abundances (blue, pink, and red) between 23.7 and 24 $\mu$m. The flux is normalized to the 9$_{8,1}$ -- 8$_{7,2}$ line at 23.87 $\mu$m.}
    \label{fig:spectra_H2O_depl}
\end{figure}

The only difference between the full chemistry and parameterized abundance models is the abundance structure. Comparing the left panels of Fig. \ref{fig:abu_fl9}, it is clear that the parameterized model has a much higher \ce{H2O} abundance above the snow surface with our choice of $n$(\ce{H2O})/$n_{\rm H}$ = $10^{-4}$. The lower \ce{H2O} abundance in the full chemistry model is due to the efficient photodissociation of \ce{H2O} in this layer, as the temperature is only high enough for efficient gas-phase formation (>300 K) in a small sliver at the surface. Thus, the \ce{H2O} column this far out in the disk cannot efficiently self-shield against destruction. Both models adopt the same temperature structure, so this cannot be the cause of the observed differences. In fact, a parameterized model with the assumption of $T_{\rm gas} = T_{\rm dust}$ (as typically done in previous work; e.g., \citealt{bruderer2015, bosman2017}) reveals only a negligible change to the strength of the cold emission (an increase of $\sim$10\%), whereas the warm and hot emission is affected much more strongly. \\
\newline
As such, we further investigate the impact of the abundance structure on the \ce{H2O} emission. We run two additional parameterized models where the same abundance structure as before is adopted, but now the \ce{H2O} abundance beyond 1 au is lowered by 1 or 2 orders of magnitude to $10^{-5}$ and $10^{-6}$, which allows the models to more closely match the abundance structure seen in the full chemistry model. The \ce{H2O} abundance structure of these new models, along with the fiducial full chemistry and parameterized models, are presented in Fig. \ref{fig:Abu_H2O_depl}. The contours represent the 70\% CBFs of the three aforementioned lines. The emitting regions of the two higher-$E_{\rm up}$ lines remain relatively unchanged between the three parameterized models, but the emitting region of the coldest line is clearly affected by the lower abundance in the outer surface layers. {The emitting region of the warm \ce{H2O} (pink contours in Figs. \ref{fig:abu_fl9} and \ref{fig:Abu_H2O_depl}) also increases between the parameterized and full chemistry models, as the parameterized abundance structure now allows it to be abundant at slightly lower temperatures than where it would otherwise form efficiently.} \\
\newline
Fig. \ref{fig:spectra_H2O_depl} clearly demonstrates the effect of lowering the \ce{H2O} abundance above the snow surface, by showing four lines of interest between 23.7 and 24 $\mu$m which are a key diagnostic for cold \ce{H2O}. Two of these lines (the 9$_{8,1}$ -- 8$_{7,2}$ and 11$_{5,6}$ -- 10$_{4,7}$ lines) have an $E_{\rm up}$ of $\sim$3000 K, whereas the other two lines (the 8$_{3,6}$ -- 7$_{0,7}$ and 8$_{4,5}$ -- 7$_{1,6}$ lines) have a lower $E_{\rm up}$ of $\sim$1500 K. {We note that the emitting region of the 8$_{3,6}$ -- 7$_{0,7}$ line is shown in Fig. \ref{fig:abu_fl9} in dark blue, and the other 1500 K line traces a nearly identical region in the disk. The two 3000 K lines trace a very similar region as the 2438 K line shown in pink in Fig. \ref{fig:abu_fl9}, though their emission is concentrated slightly higher up (similar to the vertical extent of the 1500 K lines). The flux ratio of these four lines} is very sensitive to the temperature of the gas \citep[see, e.g.,][]{banzatti2023, banzatti2025, temmink2024b}, with the 1500 K lines increasing rapidly in strength with respect to the 3000 K lines when the temperature of the gas is decreased. Additionally, \citet{banzatti2025} point out that an asymmetry between the 8$_{3,6}$ -- 7$_{0,7}$ and 8$_{4,5}$ -- 7$_{1,6}$ lines can constrain the temperature of the coldest emission even further, with a line ratio of $\gtrsim$1.2 requiring gas temperatures $\lesssim$220 K. \\
\newline
We normalize the synthetic spectra of all four models to the 9$_{8,1}$ -- 8$_{7,2}$ line between the two 1500 K lines. In the full chemistry model (shown in black), all four lines have a similar flux, whereas in the fiducial parameterized model (shown in blue), the colder $\sim$1500 K lines are stronger and show a prominent asymmetry. The strength of these cold lines with respect to the warmer 3000 K lines is reduced as the \ce{H2O} abundance above the snow surface is lowered (pink and red lines). We note that these findings do not change when LTE instead of non-LTE excitation is considered. As such, it seems that a relatively high \ce{H2O} abundance above the snow surface ($\gtrsim$10$^{-5}$) beyond $\sim$1 au is required for the spectrum to show prominent cold \ce{H2O} emission and be consistent with several observations \citep[see][]{banzatti2023, banzatti2025, temmink2024b, romero-mirza2024_sample}.\\
\newline
In Sect. \ref{app:tgas}, we demonstrate that such a high \ce{H2O} abundance above the snow surface at large radii (> 1 au) is very hard to attain through gas-phase formation alone. The temperature in this layer is simply too low, as is also reflected by the fact that such a reservoir is not present in our fiducial chemistry model (see Fig. \ref{fig:abu_fl9}). Fig. \ref{fig:abu_Tgas} demonstrates that a strong increase in gas temperature in this layer can produce abundant \ce{H2O} above the snow surface (mimicking the abundance structure in our parameterized model), however this does not come paired with a strong enhancement of the cold \ce{H2O} with respect to the hot and warm emission. Hence, this implies that the prominent cold \ce{H2O} seen in observations requires a high \ce{H2O} abundance above the snow surface combined with a low temperature $\lesssim$300 K.

\subsubsection{Temperature diagnostics}

To gain a further, comprehensive insight into the temperature distribution probed by the \ce{H2O} emission from our models, we create a diagnostic diagram following the method described in \citet{banzatti2025}. In this work, the authors define several clean, non-blended diagnostic \ce{H2O} lines with $E_{\rm up}$ of 6000, 3600 and 1500 K. Their line ratios can then provide useful insight on the relative strengths of different temperature components in the \ce{H2O} spectrum. The 3600/6000 K line ratio captures the relative strength of the warm ($\sim$400 K) component with respect to the hot ($\sim$850 K), and the 1500/3600 K provides the relative strength of the colder component ($\sim$200 K) with respect to the warm component. Plotting these two ratios against each other can then, at a glance, demonstrate what temperature component a spectrum (be it a model or an observation) is dominated by.\\
\begin{figure}
    \centering
    \includegraphics[width=\linewidth]{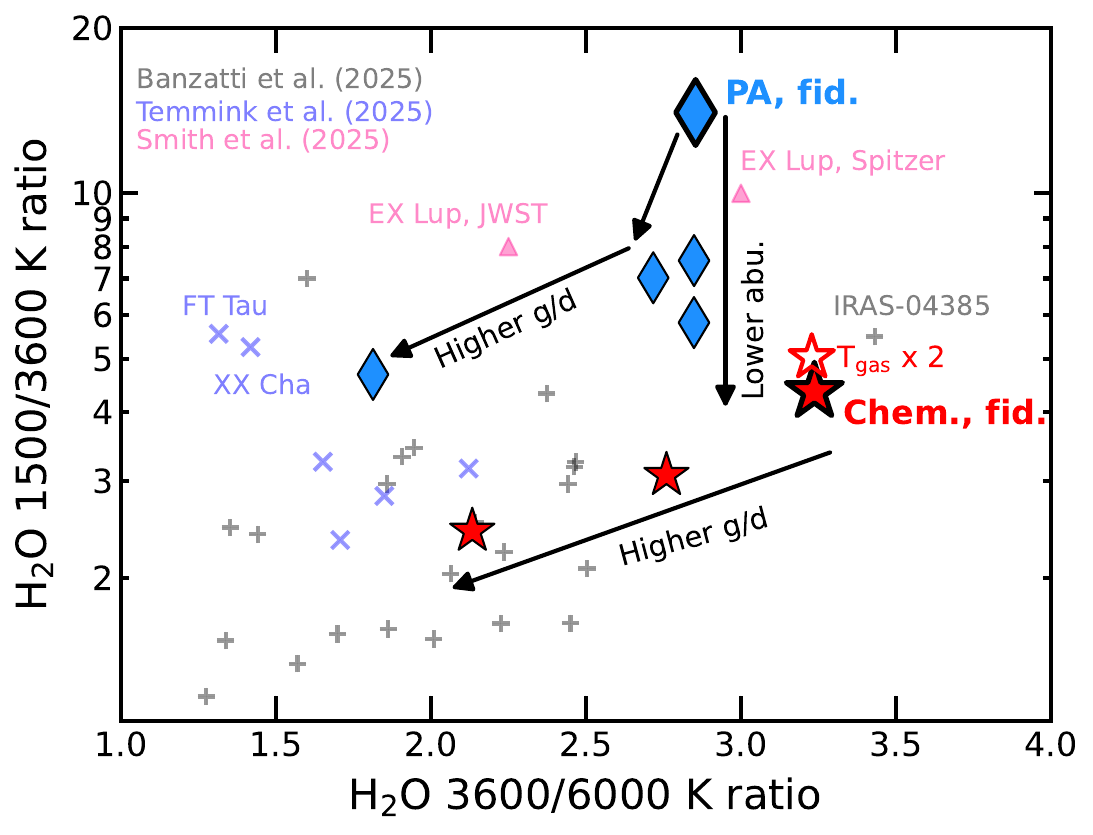}
    \caption{\ce{H2O} temperature diagnostic diagram as defined in \citet{banzatti2025}. The observations presented \citet{banzatti2025}, \citet{temmink2025}, and \citet{smith2025} are shown as grey plus symbols, blue crosses, and pink triangles, respectively. The parameterized models presented in this work are shown as blue diamonds and the chemistry models are shown as red stars. In both cases, the fiducial model is indicated as the largest symbol with a thicker black outline.}
    \label{fig:B25_plot}
\end{figure}
\newline
The temperature diagnostic diagram is presented in Fig. \ref{fig:B25_plot}, where the samples presented in \citet{banzatti2025} and \citet{temmink2025} are shown in gray plus symbols and blue crosses, respectively. EX Lup, as presented in \citet{smith2025}, is indicated using pink triangles. The parameterized abundance (PA) models are shown in blue diamonds and the chemistry models are shown in red stars. For both sets of models, the fiducial model is indicated as the largest marker with a thicker outline. Models with a higher fraction of large grains ($f_\ell$) and thus a higher gas-to-dust ratio in the IR emitting region (see also Sect. \ref{app:fl999}) are also shown. These models lie to the lower-left compared to the fiducial model, indicating that they have a stronger warm component and a weaker cold component compared to the fiducial models. This can be explained by the increased temperature in the surface layers due to the increased dust settling in these models. \\
\begin{figure*}
    \centering
    \includegraphics[width=\linewidth]{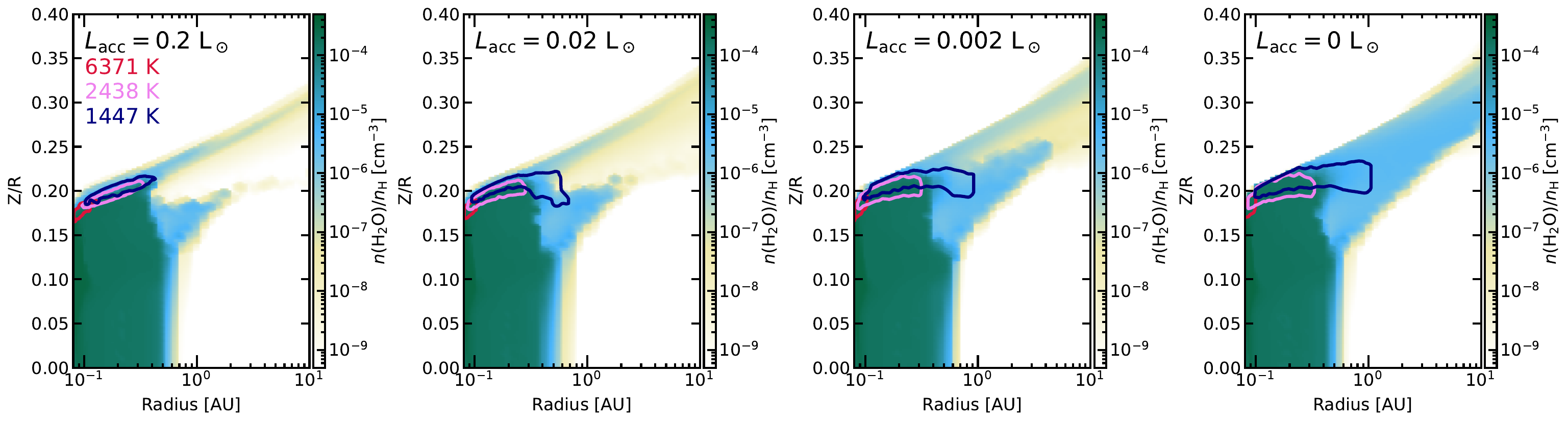}
    \caption{\ce{H2O} abundance maps of four models using the chemical network with varying accretion luminosity. The $\tau=1$ surface of the dust continuum at 15 $\mu$m is indicated with a blue dashed line. Red, pink, and blue contours represent the 70\% emitting regions of the \ce{H2O} 17$_{7,10}$ -- 16$_{4,13}$ ($E_{\rm up} = 6371 K$), 11$_{3,9}$ -- 10$_{0,10}$ ($E_{\rm up} = 2438 K$), and 8$_{3,6}$ -- 7$_{0,7}$ ($E_{\rm up} = 1447 K$) lines, respectively. }
    \label{fig:Abu_H2O_Lacc}
\end{figure*}
\newline
The two additional parameterized models presented in Fig. \ref{fig:Abu_H2O_depl} lie directly below the fiducial parameterized model in Fig. \ref{fig:B25_plot}, indicating that they have a weaker cold component, but their warm and hot components are unaffected. This demonstrates that the strength of the cold \ce{H2O} component in a spectrum is directly linked to the \ce{H2O} abundance above the snow surface. {The model with increased gas temperature (see Fig. \ref{fig:abu_Tgas}) shows only slightly colder \ce{H2O} compared to the fiducial chemistry model, and does not come close to matching the parameterized models, even though the abundance structure is similar.} We note that all models lie further to the right compared to most of the data. This could be interpreted as evidence for a high gas-to-dust ratio in the upper layers of these disks, however it is also possible that our models lack sufficient hot ($\gtrsim$800 K) \ce{H2O}. \citet{banzatti2025} demonstrate the presence of high $E_{\rm up}$ {lines} (>7200 K) throughout the rotational \ce{H2O} spectrum, which are not present in our synthetic spectra due to the $E_{\rm up}$ cutoff at 7200 K in the molecular data used {due to the lack of collisional rate coefficients for these levels \citep{faure2008}}. The inclusion of these lines may enhance the hot component and allow our models to better match the observations. \\
\newline
In Fig. \ref{fig:B25_plot}, several disks are labeled that are known to show a strong cold \ce{H2O} component in their spectra: IRAS-04385 \citep{banzatti2025}, FT Tau, and XX Cha \citep{temmink2025}. The 1500/3600 K line ratio of FT Tau and XX Cha, two disks which are noted to have strongly enhanced cold \ce{H2O}, generally agrees with our parameterized models, thus potentially indicating that these sources indeed have a relatively high \ce{H2O} abundance above the snow surface. The same holds true for EX Lup, which was also demonstrated to have strong cold \ce{H2O} by \citet{smith2025}, who speculate that this may be the result of an outburst (see also Sect. \ref{subsec:accretion}). \\
\newline
Comparing our models to IRAS-04385 brings up another interesting finding: its 1500/3600 and 3600/6000 K line ratios match our fiducial full chemistry model very closely, yet their spectra are visually quite different, especially at 23.8 $\mu$m. In IRAS-04385, the two 1500 K lines are much stronger than the 3000 K line (resembling our fiducial parameterized model, the blue line in Fig. \ref{fig:spectra_H2O_depl}; see Fig. 11 in \citealt{banzatti2025}) whereas this line ratio is relatively flat in our fiducial full chemistry model (black line in Fig. \ref{fig:spectra_H2O_depl}). This indicates that the 1500/3600 K line ratio is not always directly indicative of what the lines at 23.8 $\mu$m look like, and some caution should be taken before it is definitively concluded whether a source has strong cold \ce{H2O} emission or not.

\section{Discussion}\label{sec:discussion}

The analysis presented in Figs. \ref{fig:spectra_H2O_depl} and \ref{fig:B25_plot} indicates that a high \ce{H2O} abundance ($\gtrsim$10$^{-5}$) above the snow surface at radii $\gtrsim$1 au is required for prominent cold \ce{H2O} emission to be present in the spectrum. However, such a high abundance is not naturally present in our full chemistry model as the {temperature is too low and} \ce{H2O} is efficiently photodissociated in this region. We explore several possibilities how one can obtain a higher \ce{H2O} abundance above the snow surface, to understand how this prominent cold \ce{H2O} emission could arise in observations.

\begin{figure}
    \centering
    \includegraphics[width=\linewidth]{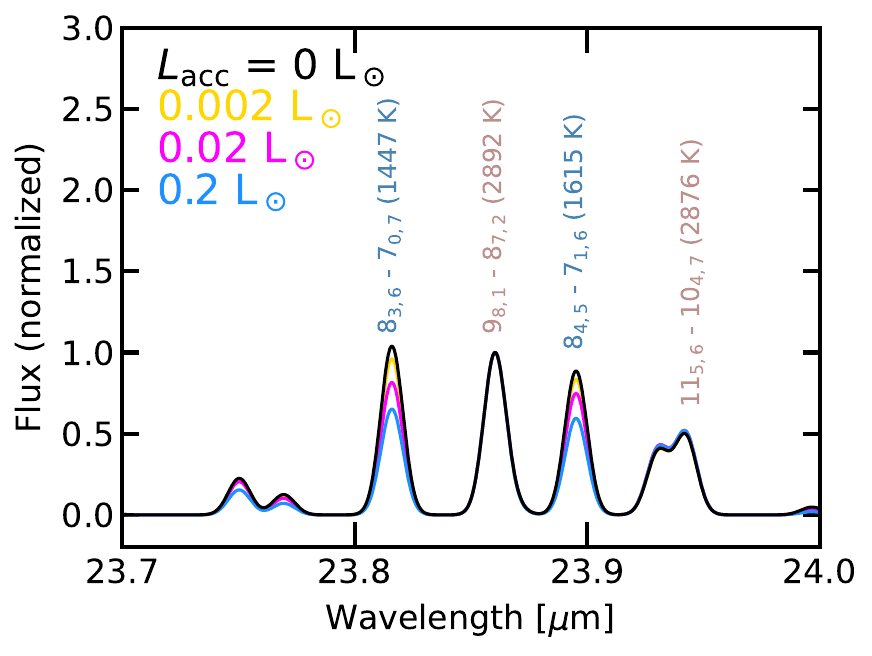}
    \caption{Left: input blackbody spectra for an $L_* = 0.8 L_\odot$ source with an accretion luminosity of $L_{\rm acc} = 0, 0.002, 0.02$ and $0.2 L_\odot$ (black, yellow, pink, and blue). Right: synthetic \ce{H2O} spectra corresponding to the four models in the left panel.}
    \label{fig:spectra_H2O_Lacc}
\end{figure}

\subsection{Lower FUV luminosity}\label{subsec:low_FUV}

First, we investigate whether a lower photodissociation rate could enhance the \ce{H2O} abundance above the snow surface, leading to stronger cold \ce{H2O} emission, for example due to a lower FUV flux coming from the central source. To this end, a small grid of four models with full chemistry and varying accretion (FUV) luminosity was run. {Instead of the AS 209 input stellar spectrum, we use} a 4000 K blackbody with an additional 10000 K component that is scaled by the mass accretion rate \citep[see][]{kama2016, visser2018}. {All other parameters related to the disk structure and input elemental abundances are kept constant. This is summarized in Table \ref{tab:params}.} The resulting \ce{H2O} abundance structure is presented in Fig. \ref{fig:Abu_H2O_Lacc}. Clearly, the \ce{H2O} abundance above the snow surface increases as the input FUV luminosity of the model decreases due to the decreased photodissociation rate. \\
\newline
The resulting synthetic spectra from the four models are presented in Fig. \ref{fig:spectra_H2O_Lacc}, once again showing the 23.7 to 24 $\mu$m region. As expected due to the increase in \ce{H2O} abundance with decreasing FUV luminosity, the cold 1500 K lines (8$_{3,6}$ -- 7$_{0,7}$ and 8$_{4,5}$ -- 7$_{1,6}$) indeed increase in strength with respect to the 3000 K lines (9$_{8,1}$ -- 8$_{7,2}$ and 11$_{5,6}$ -- 10$_{4,7}$) with decreasing FUV luminosity. However, the difference in line ratio is not nearly as apparent as in Fig. \ref{fig:spectra_H2O_depl}; the \ce{H2O} abundance above the snow surface, while it does increase when the input FUV luminosity is lowered, does not exceed $n$(\ce{H2O})/$n_{\rm H} \sim 10^{-6}$. As such, the variety in this line ratio as seen in MIRI observations \citep[see, e.g.,][]{banzatti2023, banzatti2025, romero-mirza2024_sample} cannot be explained solely by a difference in FUV luminosity. 

\subsection{Vertical mixing and enhanced \ce{H2O} self-shielding}\label{subsec:mixing}

Second, we should consider that our thermochemical models are static, and thus do not include dynamical processes such as dust transport. Recent work has shown growing evidence that dust dynamics, specifically radial drift, may play an important role in disks that show strong cold \ce{H2O} emission. These disks {with strong cold \ce{H2O}} are found to be generally more compact in millimeter dust emission than the disks in which the cold \ce{H2O} emission is less prominent \citep{banzatti2020, banzatti2023, banzatti2025, romero-mirza2024_sample}, although whether this generally holds is being challenged by new data. \citet{temmink2025} find that not all of the compact disks in their sample show an enhancement in the cold emission. It is postulated that the cold \ce{H2O} emission is a signature of strong radial drift in these disks that brings a large volume of ices across the \ce{H2O} snowline. The subsequent sublimation increases the amount of cold vapor mass at the snowline, which translates into an enhanced flux of the coldest \ce{H2O} lines. However, our work has shown that it is not the reservoir within the midplane snowline that the coldest \ce{H2O} emission is most sensitive to. Instead, it is the reservoir above the snow surface at larger radii that impacts these lines the most. As such, there needs to be a contribution from vertical mixing that carries the icy grains upwards across the 2D snow surface, in conjunction with the radial drift. \\
\begin{figure}
    \centering
    \includegraphics[width=0.8\linewidth]{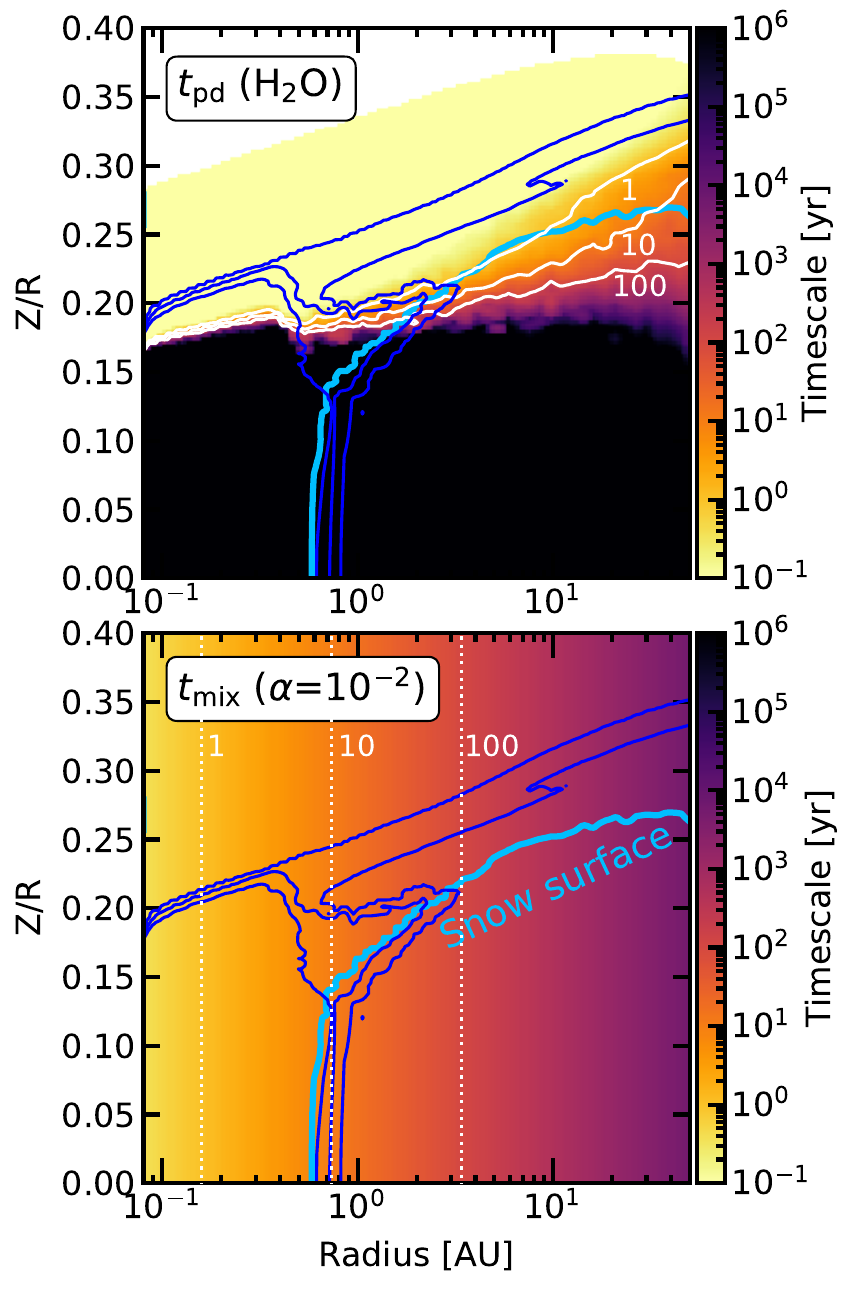}
    \caption{\ce{H2O} photodissociation timescale and turbulent diffusion mixing timescale of the fiducial model using the chemical network. The 1, 10 and 100 year timescale contours are indicated in white in both panels. {The dark blue contours in both panels indicates the $10^{-5}$, $10^{-6}$, and $10^{-8}$ \ce{H2O} abundance contours of the fiducial chemistry model (see also top left panel of Fig. \ref{fig:abu_fl9}).} The light blue line represents the \ce{H2O} snow surface. }
    \label{fig:timescales}
\end{figure}
\newline 
For vertical mixing to enhance the \ce{H2O} abundance to the degree that seems to be required from our modeling (reaching a fractional abundance of $\gtrsim$10$^{-5}$), it needs to overcome the fast photodissociation rates in the layer above the snow surface. In the left panel of Fig. \ref{fig:timescales}, we present the \ce{H2O} photodissociation timescale for the fiducial chemistry model. In the right panel, we present a simple calculation of the turbulent vertical mixing timescale, following \citet{xie1995_turbulence} and \citet{semenov2006_turbulence}. These works assume the vertical mixing to be set by turbulent diffusion, thus allowing the mixing timescale to be approximated as $t_{\rm mix} \approx H^2/D_{\rm turb}$. Here, $H = c_{\rm s}/\Omega_{\rm K}$ is the scale height of the disk and $D_{\rm turb} = \alpha c_{\rm s}^2/\Omega_{\rm K}$ is the turbulent diffusion coefficient. $c_{\rm s}$ denotes the sound speed (assumed to be isothermal), $\Omega_{\rm K}$ denotes the Keplerian angular frequency and $\alpha$ denotes the viscosity strength. We assume a strong turbulence of $\alpha=10^{-2}$ for our calculation. We stress that this is a best-case scenario, as many observations have estimated the turbulence in disks to be much lower ($\lesssim10^{-3}$; see, e.g., \citealt{flaherty2020, villenave2025}), with only a few exceptions \citep[e.g.,][]{Paneque2024, flaherty2024}.\\
\newline
Fig. \ref{fig:timescales} shows that the photodissociation timescale is short ($\lesssim$1 yr{; seen also in, e.g., \citealt{xu2019, kanwar2024_model}}) in the surface layer above the snow surface where the \ce{H2O} abundance is correlated with the cold \ce{H2O} emission. The mixing timescale, on the other hand, is much longer (10-100 yrs), even in the best-case scenario with very strong turbulence. Thus, this simple timescale comparison does not make it seem likely that vertical mixing by turbulent diffusion will be able to enhance the cold \ce{H2O} reservoir above the snow surface by a significant amount. \\
\newline
Still, a comparison of timescales alone may not provide the full picture. The effects of vertical mixing by turbulent diffusion have been implemented in the thermochemical code ProDiMo by \citet{woitke2022} by iteratively solving the reaction-diffusion equations until convergence is reached. They show that this enhances the \ce{H2O} abundance in the uppermost surface layers ($T\gtrsim1000$ K) of the disk by up to 1-2 orders of magnitude (though their models do not reach fractional abundances higher than $\sim$$10^{-6}$ at 1 au or beyond).
These models do not consider \ce{H2O} UV-shielding \citep{bethell2009}, radial mixing, or the effects of dust transport, so the continuous replenishment of ices from the outer disk could perhaps enhance these effects even further. \\
\newline
Moreover, the simple timescale comparison does not take into account the fact that dynamical processes such as radial drift or vertical mixing may build up a sufficient column of \ce{H2O} for UV-shielding to become important. This process is accounted for during the calculation of the chemistry and thermal balance in DALI \citep{bosman2022a, bosman2022b}, however DALI does not account for the extra \ce{H2O} vapor brought to the inner disk by dynamical processes. Therefore, in reality, the effects of \ce{H2O} UV-shielding may be more efficient than what is accounted for in our models, further reducing the photodissociation rate and thus allowing for a more significant build-up of cold \ce{H2O} vapor above the snow surface.\\
\newline
A simple estimate reveals that reducing the photodissociation timescale by a factor $\sim$100 by self-shielding (which would bring it a lot closer to the mixing timescale) would require a \ce{H2O} column density of $\sim$$10^{18}$ cm$^{-2}$ (taking an average photodissociation cross section for \ce{H2O} of $\sigma_{\rm av} \sim 5\times10^{-18}$ cm$^2$; \citealt{fillion2004}). This value is not unreasonable given the constraints obtained from recent observations. \citet{temmink2024b} fit the rotational \ce{H2O} spectrum of DR Tau (where the 1500 K lines at 23.8 $\mu$m are quite strong), retrieving a column density of the coldest component between log $N$ = 17.6-17.9, depending on their exact method. The column density profiles derived by \citet{romero-mirza2024_sample} also generally come close to this value for the regions in the disk where $T\lesssim300$ K. \\
\newline
Thus, it seems that a clear consensus on whether vertical mixing can produce the \ce{H2O} abundance needed to produce strong cold \ce{H2O} emission would require more sophisticated dynamical modeling. The timescales by themselves suggest that the photodissociation in the disk's surface layers is too fast to allow for a significant build-up of \ce{H2O} vapor by vertical mixing, however a more thorough exploration of disk dynamics and its effects on the IR \ce{H2O} spectrum is warranted for future work. {\citet{greenwood2019_dustevol} has shown that dust evolution by itself will enhance IR emission over time due to a decreased opacity from small micron-sized grains in the surface layers, though they do not include the delivery of volatiles to the inner disk.} Alternatively, one could also consider a mechanism like surface accretion flows that can bring small icy grains inwards across the snow surface via the surface layers, rather than being mixed upwards from the midplane \citep[e.g.,][]{bitsch2014}. 

\subsection{Accretion outbursts}\label{subsec:accretion}

Finally, \citet{smith2025} provide a promising alternative explanation that links the strength of cold \ce{H2O} and the \ce{H2O} abundance above the snow surface: accretion outbursts. The authors demonstrate that EX Lup, an M0 star with known accretion variability, has undergone a cold \ce{H2O} vapor burst during its accretion outburst in 2008. This is seen by an increase in strength of the cold, low-energy \ce{H2O} lines between two \textit{Spitzer} epochs, which the authors suggest is caused by the recession of the snowline to larger radii, which allows a large volume of ice to sublimate. A comparison with more recent JWST/MIRI observations shows that this strong cold emission is still present, suggesting long (>10 yr) re-freeze-out timescales. \\
\newline
Appendix D of \citet{smith2025} demonstrates using ProDiMo thermochemical models how the \ce{H2O} abundance structure can be altered by an accretion outburst. The recession of the snowline creates a layer of highly abundant \ce{H2O} ($n$(\ce{H2O})/$n_{\rm H} \sim 10^{-4}$) right at the snow surface. This coincides very well with our predictions in Sect. \ref{subsec:res_H2O}, where we demonstrate that an increase in \ce{H2O} abundance in this layer is directly responsible for an increase in the strength of the cold \ce{H2O} lines. Moreover, their Figure 10 demonstrates that the outburst increases the \ce{H2O} column density at large radii by up to 4 orders of magnitude, reaching values of $\sim$$10^{21}$ cm$^{-2}$, meaning that this reservoir would be able to self-shield against photodissociation quite effectively. \\
\newline
{We demonstrate a similar finding in Fig. \ref{fig:abu_Tgas}, where a strong increase in gas temperature (perhaps due to an outburst) indeed leads to an enhancement in the \ce{H2O} abundance above the snow surface, resembling our fiducial parameterized model and the model presented in \citet{smith2025}. In our case, this is purely caused by the enhancement in gas-phase \ce{H2O} formation due to the increased temperature.} In our own model grid with varying accretion luminosity (Sect. \ref{subsec:low_FUV}), we assumed the accretion state to be static and thus evolved the chemistry for 1 Myr (during which the surface layers already reach a steady-state chemistry). Hence, we did not find the accretion luminosity to have such a drastic effect on the \ce{H2O} abundance on longer timescales. In contrast, \citet{smith2025} evolve their outburst model, starting from a quiescent state, for only a very short amount of time (half a year), causing the strong change in the \ce{H2O} abundance structure. This demonstrates that processes occurring on short timescales may also leave an imprint in the IR \ce{H2O} spectrum. \\
\begin{figure*}
    \centering
    \includegraphics[width=\linewidth]{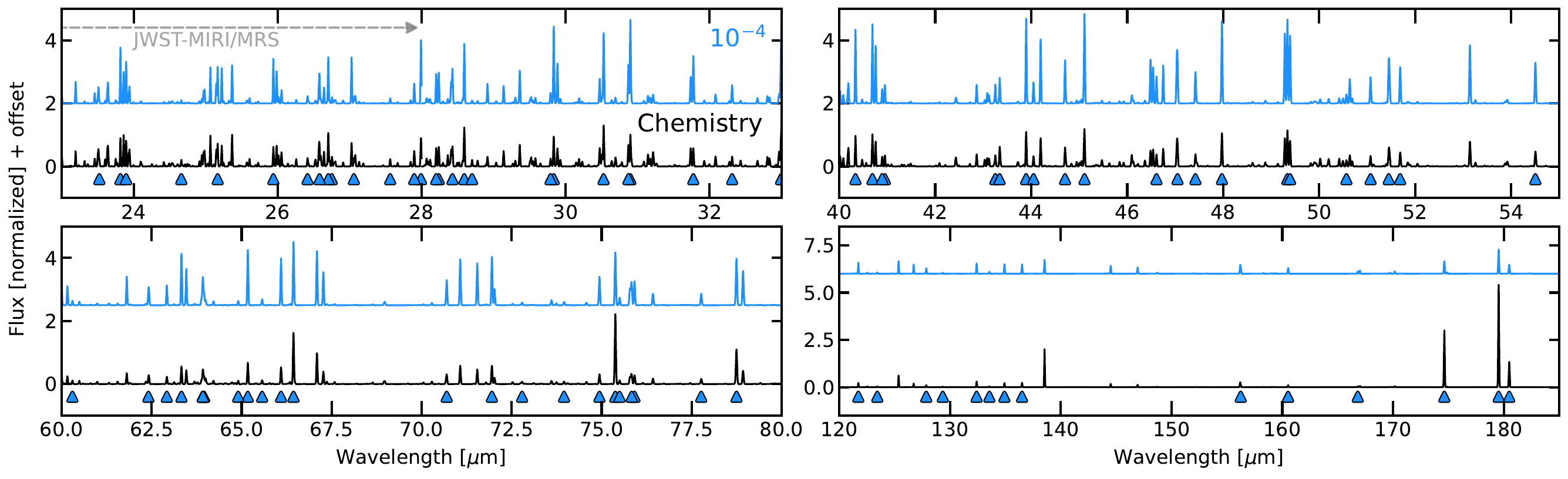}
    \caption{Synthetic \ce{H2O} spectra of the fiducial model using the chemical network (black) and the fiducial parameterized model (abundance above the snow surface of $10^{-4}$; blue) between 23 and 33, 40 and 55, 60 and 80, and 120 and 185 $\mu$m. The flux is normalized to the 9$_{8,1}$ -- 8$_{7,2}$ line (2892 K) at 23.87 $\mu$m. Blue triangles indicate \ce{H2O} lines with $E_{\rm up} < 2500$ K.}
    \label{fig:spectra_FIR}
\end{figure*}
\newline
Sect. \ref{app:tgas} further demonstrates that the low temperature in the layer above the snow surface in the fiducial chemistry model does not allow for \ce{H2O} to form abundantly (i.e. reaching $n$(\ce{H2O})/$n_{\rm H} \sim 10^{-4}$; see Fig. \ref{fig:nH2O}). This only becomes more feasible if the temperature is increased (e.g. due to an accretion outburst). However, since the latter also enhances the emission from the hot and warm \ce{H2O}, this does not lead to a spectrum dominated by cold \ce{H2O}. Instead, this implies one of two conclusions: 1) the \ce{H2O} needs to remain abundant in this layer while the disk cools down again (potentially aided by UV shielding), or 2) the \ce{H2O} needs to become abundant through a mechanism other than gas-phase formation, such as \ce{H2O} delivery from radial drift followed by vertical mixing. Given the rapid photodissociation timescales in the surface, the latter case possibly implies that some sort of continuous replenishment is required. \\
\newline
Thus, accretion outbursts seem to be able to produce a temporary burst of cold \ce{H2O} vapor that can be traced in the IR with, for example, JWST/MIRI. However, whether this explanation is generally applicable to all disks that show cold \ce{H2O} remains up for debate, as certainly not all disks show strong evidence of accretion variability. Moreover, it seems that a rather large outburst is required, as {\citet{smith2025}} did not find the moderate outburst {of EX Lup} in 2023 to cause a significant change in the spectrum.  

\subsection{Future outlook in the far-IR}

Current observations and theoretical efforts focus on the IR \ce{H2O} spectrum of disks as seen with JWST/MIRI and how dynamical processes may affect this \citep[e.g.,][]{banzatti2023, banzatti2025, romero-mirza2024_sample, kalyaan2021, kalyaan2023, mah2023}. While JWST has been able to provide many new insights into the warm ($\sim$few 100 K) \ce{H2O} content of disks, lower $E_{\rm up}$ lines tracing the colder content are only scarcely available in the spectrum, with no lines with $E_{\rm up}<800$ K present in the MIRI wavelength range. Additionally, these lines are located mostly in the longest wavelength channel which has the lowest sensitivity. To fully understand the cold \ce{H2O} content in disks, the material that is transported across the snowline by radial drift, observations at longer wavelengths, in the far-IR, are crucial \citep[see also, e.g.,][]{zhang2013, pontoppidan2018, kamp2021}.\\
\newline
While the far-IR window is currently unavailable for disk spectroscopy, the future looks promising, with both a confirmed balloon mission and a space mission under development that will be able to survey the cold \ce{H2O} content in planet-forming disks. The Planetary Origins and Evolution Multispectral Monochromator (POEMM) stratospheric balloon mission is a 2 m telescope with a high-resolution ($R=10^5$) spectrograph covering wavelengths from 35-112 $\mu$m \citep{poemm_spie}. One of its main science goals is to characterize the \ce{H2O} emission from Herbig and T Tauri disks, to further our understanding of planet formation. The PRobe far-Infrared Mission for Astrophysics (PRIMA) mission is a 1.8 m space telescope, with the FIRESS instrument providing spectroscopy of the 24-235 $\mu$m range at a superior spectral resolution to JWST/MIRI ($R=4400\times 112$ $\mu$m/$\lambda$), and with a better sensitivity than POEMM. It builds technically and scientifically on earlier design concepts such as the SPace Infrared telescope for Cosmology and Astrophysics (SPICA) mission \citep{roelfsema2018, kamp2021}.\\
\newline
Fig. \ref{fig:spectra_FIR} presents the synthetic spectra for the fiducial chemistry and parameterized models (black and blue lines). The fluxes are normalized to the flux of the 9$_{8,1}$ -- 8$_{7,2}$ line at 23.87 $\mu$m as also done in Fig. \ref{fig:spectra_H2O_depl}, but now shown from 23 to 185 $\mu$m. This illustrates that there are many more lines in the far-IR that are very sensitive to the cold \ce{H2O} abundance above the snow surface. As such, these lines may also make for good tracers of the amount of \ce{H2O} ice crossing the snowline. Additionally, the two missions highlighted above will also provide much higher spectral resolution (PRIMA reaches $R\sim20\,000$ at $\lambda = 25$ $\mu$m), and can thus provide kinematical information as well for the brightest sources, allowing for the spatial distribution of the emission, and thus the location of the snowline, to be constrained.

\section{Conclusions} \label{sec:conclusions}

In this work, we have tested several retrieval techniques that are commonly used to interpret JWST observations of \ce{H2O} on synthetic DALI spectra. We summarize our findings from these tests as follows:\\
\begin{itemize}
    \item Single-temperature slab fits generally trace the warm ($\sim$500 K) component of the \ce{H2O} emission, even when the fits are performed in different narrow windows within the MIRI wavelength range. This indicates that, even in a relatively narrow window, a single temperature is not enough to capture the full complexity of the \ce{H2O} emission.
    \item A three-component MCMC fit to the entire MIRI wavelength range (following, e.g., \citealt{temmink2024b}) is better suited to capture the full gradient in temperature present in the \ce{H2O} spectrum, also capturing the contributions from hotter (>600 K) and colder (<300 K) emission. 
    \item Retrieved temperatures generally agree well with the model, though they may slightly underestimate the true temperature in the emitting layers due to non-LTE effects such as sub-thermal excitation. 
    \item The retrieved column density, both from single- and three-component fits, generally traces close to the dust $\tau=1$ surface for our fiducial models. However, in models with more settled dust where the $\tau=1$ surface lies deeper into the disk, the lines become optically thick before the dust does and the emission therefore does not trace all the way down to the $\tau=1$ surface.
    \item The retrievals on \ce{CO2} emission find the same conclusions as above, and we find that the \ce{^13CO2} emission retrieves a lower temperature than \ce{^12CO2} due to it tracing deeper into the disk.
\end{itemize}
Additionally, we further investigate the cold ($\sim$200 K) component of the \ce{H2O} emission to understand the origin of this emission. We summarize our findings as follows:
\begin{itemize}
    \item We demonstrate that the strength of the cold \ce{H2O} emission is directly linked to the \ce{H2O} abundance above the snow surface at larger radii ($\gtrsim$1 au). This implies that the disks in which strong cold \ce{H2O} has been observed \citep[e.g.,][]{banzatti2023, banzatti2025} likely have a high abundance ($\gtrsim10^{-5}$) in this reservoir, higher than is predicted by our fiducial full chemistry model. 
    \item {Additionally, we demonstrate that the temperature needs to be low (<300 K), meaning this high abundance must be achieved through a mechanism other than gas-phase formation, which is not efficient at these temperatures in gas exposed to intense UV radiation.}    
    \item A lower FUV luminosity from the central star, for example due to a low mass accretion rate, can increase the \ce{H2O} abundance above the snow surface. However, this effect is too small to explain the diversity in cold \ce{H2O} strength seen in observations.
    \item Instead, vertical mixing of ices across the snow surface, in conjunction with their inwards transport by radial drift, could possibly enhance the \ce{H2O} abundance beyond what is considered in our static thermochemical models. This would fall in line with recent work tentatively identifying cold \ce{H2O} emission as a signature of strong radial drift \citep{banzatti2023, banzatti2025, romero-mirza2024_sample}, though further investigation is needed to confirm this. 
    \item Accretion outbursts provide an interesting alternative explanation for enhanced cold \ce{H2O}, as \citet{smith2025} demonstrate. These outbursts may enhance the \ce{H2O} abundance above the snow surface, which our work has shown directly correlates with the strength of cold emission.
\end{itemize}
With this work, we have provided further indications from a modeling perspective that the cold \ce{H2O} emission seen with JWST may be a signature of a dynamical process, as argued previously from an observational perspective. To understand the origins of the cold \ce{H2O} emission, and to give us further insights into how JWST observations could possibly probe the \ce{H2O} at its snowline, the effects of dust transport and mixing on the observed IR emission should be investigated in future work. 

\begin{acknowledgements}
    We thank the referee for their constructive feedback that helped improve this manuscript. Astrochemistry in Leiden is supported by funding from the European Research Council (ERC) under the European Union’s Horizon 2020 research and innovation programme (grant agreement No. 101019751 MOLDISK). E.v.D. also acknowledges support from the Danish National Research Foundation through the Center of Excellence ``InterCat'' (DNRF150).
\end{acknowledgements}

\bibliographystyle{aa}
\bibliography{references}

\begin{thebibliography}{81}
\expandafter\ifx\csname natexlab\endcsname\relax\def\natexlab#1{#1}\fi

\bibitem[{{Allen} {et~al.}(1980){Allen}, {Scragg}, \& {Simpson}}]{allen1980}
{Allen}, D., {Scragg}, T., \& {Simpson}, C. 1980, Chemical Physics, 51, 279

\bibitem[{{Arulanantham} {et~al.}(2025){Arulanantham}, {Salyk}, {Pontoppidan}, {Banzatti}, {Zhang}, {{\"O}berg}, {Long}, {Carr}, {Najita}, {Pascucci}, {Colmenares}, {Xie}, {Huang}, {Green}, {Andrews}, {Blake}, {Bergin}, {Pinilla}, {Vioque}, {Dahl}, {Raul}, {Krijt}, \& {The Jdiscs Collaboration}}]{arulanantham2025}
{Arulanantham}, N., {Salyk}, C., {Pontoppidan}, K., {et~al.} 2025, \aj, 170, 67

\bibitem[{{Banzatti} {et~al.}(2020){Banzatti}, {Pascucci}, {Bosman}, {Pinilla}, {Salyk}, {Herczeg}, {Pontoppidan}, {Vazquez}, {Watkins}, {Krijt}, {Hendler}, \& {Long}}]{banzatti2020}
{Banzatti}, A., {Pascucci}, I., {Bosman}, A.~D., {et~al.} 2020, \apj, 903, 124

\bibitem[{{Banzatti} {et~al.}(2023){Banzatti}, {Pontoppidan}, {Carr}, {Jellison}, {Pascucci}, {Najita}, {Mu{\~n}oz-Romero}, {{\"O}berg}, {Kalyaan}, {Pinilla}, {Krijt}, {Long}, {Lambrechts}, {Rosotti}, {Herczeg}, {Salyk}, {Zhang}, {Bergin}, {Ballering}, {Meyer}, {Bruderer}, \& {Jdiscs Collaboration}}]{banzatti2023}
{Banzatti}, A., {Pontoppidan}, K.~M., {Carr}, J.~S., {et~al.} 2023, \apjl, 957, L22

\bibitem[{{Banzatti} {et~al.}(2025){Banzatti}, {Salyk}, {Pontoppidan}, {Carr}, {Zhang}, {Arulanantham}, {Krijt}, {{\"O}berg}, {Cleeves}, {Najita}, {Pascucci}, {Blake}, {Romero-Mirza}, {Bergin}, {Cieza}, {Pinilla}, {Long}, {Mallaney}, {Xie}, {Waggoner}, {Kaeufer}, \& {The Jdiscs Collaboration}}]{banzatti2025}
{Banzatti}, A., {Salyk}, C., {Pontoppidan}, K.~M., {et~al.} 2025, \aj, 169, 165

\bibitem[{{Barber} {et~al.}(2006){Barber}, {Tennyson}, {Harris}, \& {Tolchenov}}]{barber2006}
{Barber}, R.~J., {Tennyson}, J., {Harris}, G.~J., \& {Tolchenov}, R.~N. 2006, \mnras, 368, 1087

\bibitem[{{Bethell} \& {Bergin}(2009)}]{bethell2009}
{Bethell}, T. \& {Bergin}, E. 2009, Science, 326, 1675

\bibitem[{{Bitsch} {et~al.}(2014){Bitsch}, {Morbidelli}, {Lega}, {Kretke}, \& {Crida}}]{bitsch2014}
{Bitsch}, B., {Morbidelli}, A., {Lega}, E., {Kretke}, K., \& {Crida}, A. 2014, \aap, 570, A75

\bibitem[{{Bosman} {et~al.}(2022{\natexlab{a}}){Bosman}, {Bergin}, {Calahan}, \& {Duval}}]{bosman2022a}
{Bosman}, A.~D., {Bergin}, E.~A., {Calahan}, J., \& {Duval}, S.~E. 2022{\natexlab{a}}, \apjl, 930, L26

\bibitem[{{Bosman} {et~al.}(2022{\natexlab{b}}){Bosman}, {Bergin}, {Calahan}, \& {Duval}}]{bosman2022b}
{Bosman}, A.~D., {Bergin}, E.~A., {Calahan}, J.~K., \& {Duval}, S.~E. 2022{\natexlab{b}}, \apjl, 933, L40

\bibitem[{{Bosman} {et~al.}(2017){Bosman}, {Bruderer}, \& {van Dishoeck}}]{bosman2017}
{Bosman}, A.~D., {Bruderer}, S., \& {van Dishoeck}, E.~F. 2017, \aap, 601, A36

\bibitem[{{Bruderer}(2013)}]{bruderer2013}
{Bruderer}, S. 2013, \aap, 559, A46

\bibitem[{{Bruderer} {et~al.}(2015){Bruderer}, {Harsono}, \& {van Dishoeck}}]{bruderer2015}
{Bruderer}, S., {Harsono}, D., \& {van Dishoeck}, E.~F. 2015, \aap, 575, A94

\bibitem[{{Bruderer} {et~al.}(2012){Bruderer}, {van Dishoeck}, {Doty}, \& {Herczeg}}]{bruderer2012}
{Bruderer}, S., {van Dishoeck}, E.~F., {Doty}, S.~D., \& {Herczeg}, G.~J. 2012, \aap, 541, A91

\bibitem[{{Charnley}(1997)}]{charnley1997}
{Charnley}, S.~B. 1997, \apj, 481, 396

\bibitem[{{Dawson} \& {Johnson}(2018)}]{dawson2018}
{Dawson}, R.~I. \& {Johnson}, J.~A. 2018, \araa, 56, 175

\bibitem[{{Faure} \& {Josselin}(2008)}]{faure2008}
{Faure}, A. \& {Josselin}, E. 2008, \aap, 492, 257

\bibitem[{{Fillion} {et~al.}(2004){Fillion}, {Ruiz}, {Yang}, {Castillejo}, {Rostas}, \& {Lemaire}}]{fillion2004}
{Fillion}, J.~H., {Ruiz}, J., {Yang}, X.~F., {et~al.} 2004, \jcp, 120, 6531

\bibitem[{{Flaherty} {et~al.}(2020){Flaherty}, {Hughes}, {Simon}, {Qi}, {Bai}, {Bulatek}, {Andrews}, {Wilner}, \& {K{\'o}sp{\'a}l}}]{flaherty2020}
{Flaherty}, K., {Hughes}, A.~M., {Simon}, J.~B., {et~al.} 2020, \apj, 895, 109

\bibitem[{{Flaherty} {et~al.}(2024){Flaherty}, {Hughes}, {Simon}, {Reina}, {Qi}, {Bai}, {Andrews}, {Wilner}, \& {K{\'o}sp{\'a}l}}]{flaherty2024}
{Flaherty}, K., {Hughes}, A.~M., {Simon}, J.~B., {et~al.} 2024, \mnras, 532, 363

\bibitem[{{Foreman-Mackey} {et~al.}(2013){Foreman-Mackey}, {Hogg}, {Lang}, \& {Goodman}}]{foreman_2013_emcee}
{Foreman-Mackey}, D., {Hogg}, D.~W., {Lang}, D., \& {Goodman}, J. 2013, \pasp, 125, 306

\bibitem[{{Gasman} {et~al.}(2025){Gasman}, {Temmink}, {van Dishoeck}, {Kurtovic}, {Grant}, {Sellek}, {Tabone}, {Henning}, {Kamp}, {G{\"u}del}, {Barrado}, {Caratti o Garatti}, {Glauser}, {Waters}, {Arabhavi}, {Jang}, {Kanwar}, {Lienert}, {Perotti}, {Schwarz}, \& {Vlasblom}}]{gasman2025}
{Gasman}, D., {Temmink}, M., {van Dishoeck}, E.~F., {et~al.} 2025, \aap, 694, A147

\bibitem[{{Gasman} {et~al.}(2023){Gasman}, {van Dishoeck}, {Grant}, {Temmink}, {Tabone}, {Henning}, {Kamp}, {G{\"u}del}, {Lagage}, {Perotti}, {Christiaens}, {Samland}, {Arabhavi}, {Argyriou}, {Abergel}, {Absil}, {Barrado}, {Boccaletti}, {Bouwman}, {Caratti o Garatti}, {Geers}, {Glauser}, {Guadarrama}, {Jang}, {Kanwar}, {Lahuis}, {Morales-Calder{\'o}n}, {Mueller}, {Nehm{\'e}}, {Olofsson}, {Pantin}, {Pawellek}, {Ray}, {Rodgers-Lee}, {Scheithauer}, {Schreiber}, {Schwarz}, {Vandenbussche}, {Vlasblom}, {Waters}, {Wright}, {Colina}, {Greve}, \& {{\"O}stlin}}]{gasman2023b}
{Gasman}, D., {van Dishoeck}, E.~F., {Grant}, S.~L., {et~al.} 2023, \aap, 679, A117

\bibitem[{{Grant} {et~al.}(2023){Grant}, {van Dishoeck}, {Tabone}, {Gasman}, {Henning}, {Kamp}, {G{\"u}del}, {Lagage}, {Bettoni}, {Perotti}, {Christiaens}, {Samland}, {Arabhavi}, {Argyriou}, {Abergel}, {Absil}, {Barrado}, {Boccaletti}, {Bouwman}, {o Garatti}, {Geers}, {Glauser}, {Guadarrama}, {Jang}, {Kanwar}, {Lahuis}, {Morales-Calder{\'o}n}, {Mueller}, {Nehm{\'e}}, {Olofsson}, {Pantin}, {Pawellek}, {Ray}, {Rodgers-Lee}, {Scheithauer}, {Schreiber}, {Schwarz}, {Temmink}, {Vandenbussche}, {Vlasblom}, {Waters}, {Wright}, {Colina}, {Greve}, {Justannont}, \& {{\"O}stlin}}]{grant2023}
{Grant}, S.~L., {van Dishoeck}, E.~F., {Tabone}, B., {et~al.} 2023, \apjl, 947, L6

\bibitem[{{Greenwood} {et~al.}(2019){Greenwood}, {Kamp}, {Waters}, {Woitke}, \& {Thi}}]{greenwood2019_dustevol}
{Greenwood}, A.~J., {Kamp}, I., {Waters}, L.~B.~F.~M., {Woitke}, P., \& {Thi}, W.~F. 2019, \aap, 626, A6

\bibitem[{{Henning} {et~al.}(2024){Henning}, {Kamp}, {Samland}, {Arabhavi}, {Kanwar}, {van Dishoeck}, {G{\"u}del}, {Lagage}, {Waelkens}, {Abergel}, {Absil}, {Barrado}, {Boccaletti}, {Bouwman}, {Caratti o Garatti}, {Geers}, {Glauser}, {Lahuis}, {Mueller}, {Nehm{\'e}}, {Olofsson}, {Pantin}, {Ray}, {Scheithauer}, {Vandenbussche}, {Waters}, {Wright}, {Argyriou}, {Christiaens}, {Franceschi}, {Gasman}, {Grant}, {Guadarrama}, {Jang}, {Morales-Calder{\'o}n}, {Pawellek}, {Perotti}, {Rodgers-Lee}, {Schreiber}, {Schwarz}, {Tabone}, {Temmink}, {Vlasblom}, {Colina}, {Greve}, \& {{\"O}stlin}}]{henning2024}
{Henning}, T., {Kamp}, I., {Samland}, M., {et~al.} 2024, \pasp, 136, 054302

\bibitem[{{Hogerheijde} {et~al.}(2011){Hogerheijde}, {Bergin}, {Brinch}, {Cleeves}, {Fogel}, {Blake}, {Dominik}, {Lis}, {Melnick}, {Neufeld}, {Pani{\'c}}, {Pearson}, {Kristensen}, {Y{\i}ld{\i}z}, \& {van Dishoeck}}]{hogerheijde2011}
{Hogerheijde}, M.~R., {Bergin}, E.~A., {Brinch}, C., {et~al.} 2011, Science, 334, 338

\bibitem[{{Houge} {et~al.}(2025){Houge}, {Krijt}, {Banzatti}, {Blake}, {Pinilla}, {Pontoppidan}, {Trapman}, {Williams}, \& {Zhang}}]{houge2025}
{Houge}, A., {Krijt}, S., {Banzatti}, A., {et~al.} 2025, \mnras, 537, 691

\bibitem[{{Jacobs} {et~al.}(1975){Jacobs}, {Pettipiece}, \& {Thomas}}]{jacobs1975}
{Jacobs}, R., {Pettipiece}, K., \& {Thomas}, S. 1975, Physical Review. A, 11, 54

\bibitem[{{Kaeufer} {et~al.}(2024){Kaeufer}, {Min}, {Woitke}, {Kamp}, \& {Arabhavi}}]{kaufer2024}
{Kaeufer}, T., {Min}, M., {Woitke}, P., {Kamp}, I., \& {Arabhavi}, A.~M. 2024, \aap, 687, A209

\bibitem[{{Kalyaan} {et~al.}(2023){Kalyaan}, {Pinilla}, {Krijt}, {Banzatti}, {Rosotti}, {Mulders}, {Lambrechts}, {Long}, \& {Herczeg}}]{kalyaan2023}
{Kalyaan}, A., {Pinilla}, P., {Krijt}, S., {et~al.} 2023, \apj, 954, 66

\bibitem[{{Kalyaan} {et~al.}(2021){Kalyaan}, {Pinilla}, {Krijt}, {Mulders}, \& {Banzatti}}]{kalyaan2021}
{Kalyaan}, A., {Pinilla}, P., {Krijt}, S., {Mulders}, G.~D., \& {Banzatti}, A. 2021, \apj, 921, 84

\bibitem[{{Kama} {et~al.}(2016){Kama}, {Bruderer}, {van Dishoeck}, {Hogerheijde}, {Folsom}, {Miotello}, {Fedele}, {Belloche}, {G{\"u}sten}, \& {Wyrowski}}]{kama2016}
{Kama}, M., {Bruderer}, S., {van Dishoeck}, E.~F., {et~al.} 2016, \aap, 592, A83

\bibitem[{{Kamp} {et~al.}(2023){Kamp}, {Henning}, {Arabhavi}, {Bettoni}, {Christiaens}, {Gasman}, {Grant}, {Morales-Calder{\'o}n}, {Tabone}, {Abergel}, {Absil}, {Argyriou}, {Barrado}, {Boccaletti}, {Bouwman}, {Caratti o Garatti}, {van Dishoeck}, {Geers}, {Glauser}, {G{\"u}del}, {Guadarrama}, {Jang}, {Kanwar}, {Lagage}, {Lahuis}, {Mueller}, {Nehm{\'e}}, {Olofsson}, {Pantin}, {Pawellek}, {Perotti}, {Ray}, {Rodgers-Lee}, {Samland}, {Scheithauer}, {Schreiber}, {Schwarz}, {Temmink}, {Vandenbussche}, {Vlasblom}, {Waelkens}, {Waters}, \& {Wright}}]{kamp2023}
{Kamp}, I., {Henning}, T., {Arabhavi}, A.~M., {et~al.} 2023, Faraday Discussions, 245, 112

\bibitem[{{Kamp} {et~al.}(2021){Kamp}, {Honda}, {Nomura}, {Audard}, {Fedele}, {Waters}, {Aikawa}, {Banzatti}, {Bowey}, {Bradford}, {Dominik}, {Furuya}, {Habart}, {Ishihara}, {Johnstone}, {Kennedy}, {Kim}, {Kral}, {Lai}, {Larsson}, {McClure}, {Miotello}, {Momose}, {Nakagawa}, {Naylor}, {Nisini}, {Notsu}, {Onaka}, {Pantin}, {Podio}, {Riviere Marichalar}, {Rocha}, {Roelfsema}, {Shimonishi}, {Tang}, {Takami}, {Tazaki}, {Wolf}, {Wyatt}, \& {Ysard}}]{kamp2021}
{Kamp}, I., {Honda}, M., {Nomura}, H., {et~al.} 2021, \pasa, 38, e055

\bibitem[{{Kanwar} {et~al.}(2024){Kanwar}, {Kamp}, {Woitke}, {Rab}, {Thi}, \& {Min}}]{kanwar2024_model}
{Kanwar}, J., {Kamp}, I., {Woitke}, P., {et~al.} 2024, \aap, 681, A22

\bibitem[{{Krijt} {et~al.}(2023){Krijt}, {Kama}, {McClure}, {Teske}, {Bergin}, {Shorttle}, {Walsh}, \& {Raymond}}]{krijt2023_PPVII}
{Krijt}, S., {Kama}, M., {McClure}, M., {et~al.} 2023, in Astronomical Society of the Pacific Conference Series, Vol. 534, Protostars and Planets VII, ed. S.~{Inutsuka}, Y.~{Aikawa}, T.~{Muto}, K.~{Tomida}, \& M.~{Tamura}, 1031

\bibitem[{{Kristensen} {et~al.}(2017){Kristensen}, {van Dishoeck}, {Mottram}, {Karska}, {Y{\i}ld{\i}z}, {Bergin}, {Bjerkeli}, {Cabrit}, {Doty}, {Evans}, {Gusdorf}, {Harsono}, {Herczeg}, {Johnstone}, {J{\o}rgensen}, {van Kempen}, {Lee}, {Maret}, {Tafalla}, {Visser}, \& {Wampfler}}]{kristensen2017}
{Kristensen}, L.~E., {van Dishoeck}, E.~F., {Mottram}, J.~C., {et~al.} 2017, \aap, 605, A93

\bibitem[{{Mah} {et~al.}(2023){Mah}, {Bitsch}, {Pascucci}, \& {Henning}}]{mah2023}
{Mah}, J., {Bitsch}, B., {Pascucci}, I., \& {Henning}, T. 2023, \aap, 677, L7

\bibitem[{{Mathis} {et~al.}(1977){Mathis}, {Rumpl}, \& {Nordsieck}}]{mathis1977}
{Mathis}, J.~S., {Rumpl}, W., \& {Nordsieck}, K.~H. 1977, \apj, 217, 425

\bibitem[{{Meijerink} {et~al.}(2009){Meijerink}, {Pontoppidan}, {Blake}, {Poelman}, \& {Dullemond}}]{meijerink2009}
{Meijerink}, R., {Pontoppidan}, K.~M., {Blake}, G.~A., {Poelman}, D.~R., \& {Dullemond}, C.~P. 2009, \apj, 704, 1471

\bibitem[{{Millar} {et~al.}(2024){Millar}, {Walsh}, {Van de Sande}, \& {Markwick}}]{millar2024}
{Millar}, T.~J., {Walsh}, C., {Van de Sande}, M., \& {Markwick}, A.~J. 2024, \aap, 682, A109

\bibitem[{{Miotello} {et~al.}(2014){Miotello}, {Bruderer}, \& {van Dishoeck}}]{miotello2014}
{Miotello}, A., {Bruderer}, S., \& {van Dishoeck}, E.~F. 2014, \aap, 572, A96

\bibitem[{{Miotello} {et~al.}(2016){Miotello}, {van Dishoeck}, {Kama}, \& {Bruderer}}]{miotello2016}
{Miotello}, A., {van Dishoeck}, E.~F., {Kama}, M., \& {Bruderer}, S. 2016, \aap, 594, A85

\bibitem[{{Nevdakh} {et~al.}(2003){Nevdakh}, {Orlov}, \& {Leshenyuk}}]{nevdakh2003}
{Nevdakh}, V.~V., {Orlov}, L.~N., \& {Leshenyuk}, N.~S. 2003, Journal of Applied Spectroscopy, 70, 276

\bibitem[{{{\"O}berg} \& {Bergin}(2021)}]{oberg2021}
{{\"O}berg}, K.~I. \& {Bergin}, E.~A. 2021, \physrep, 893, 1

\bibitem[{{Paneque-Carre{\~n}o} {et~al.}(2024){Paneque-Carre{\~n}o}, {Izquierdo}, {Teague}, {Miotello}, {Bergin}, {Loomis}, \& {van Dishoeck}}]{Paneque2024}
{Paneque-Carre{\~n}o}, T., {Izquierdo}, A.~F., {Teague}, R., {et~al.} 2024, \aap, 684, A174

\bibitem[{{Pontoppidan} {et~al.}(2018){Pontoppidan}, {Bergin}, {Melnick}, {Bradford}, {Staguhn}, {Leisawitz}, {Meixner}, {Fortney}, {Salyk}, {Blake}, {Zhang}, {Banzatti}, {Kataria}, {Meshkat}, {de Val-Borro}, {Stevenson}, \& {Fraine}}]{pontoppidan2018}
{Pontoppidan}, K.~M., {Bergin}, E.~A., {Melnick}, G., {et~al.} 2018, arXiv e-prints, arXiv:1804.00743

\bibitem[{{Pontoppidan} {et~al.}(2024){Pontoppidan}, {Salyk}, {Banzatti}, {Zhang}, {Pascucci}, {{\"O}berg}, {Long}, {Mu{\~n}oz-Romero}, {Carr}, {Najita}, {Blake}, {Arulanantham}, {Andrews}, {Ballering}, {Bergin}, {Calahan}, {Cobb}, {Colmenares}, {Dickson-Vandervelde}, {Dignan}, {Green}, {Heretz}, {Herczeg}, {Kalyaan}, {Krijt}, {Pauly}, {Pinilla}, {Trapman}, \& {Xie}}]{pontoppidan2024}
{Pontoppidan}, K.~M., {Salyk}, C., {Banzatti}, A., {et~al.} 2024, \apj, 963, 158

\bibitem[{{Roelfsema} {et~al.}(2018){Roelfsema}, {Shibai}, {Armus}, {Arrazola}, {Audard}, {Audley}, {Bradford}, {Charles}, {Dieleman}, {Doi}, {Duband}, {Eggens}, {Evers}, {Funaki}, {Gao}, {Giard}, {di Giorgio}, {Gonz{\'a}lez Fern{\'a}ndez}, {Griffin}, {Helmich}, {Hijmering}, {Huisman}, {Ishihara}, {Isobe}, {Jackson}, {Jacobs}, {Jellema}, {Kamp}, {Kaneda}, {Kawada}, {Kemper}, {Kerschbaum}, {Khosropanah}, {Kohno}, {Kooijman}, {Krause}, {van der Kuur}, {Kwon}, {Laauwen}, {de Lange}, {Larsson}, {van Loon}, {Madden}, {Matsuhara}, {Najarro}, {Nakagawa}, {Naylor}, {Ogawa}, {Onaka}, {Oyabu}, {Poglitsch}, {Reveret}, {Rodriguez}, {Spinoglio}, {Sakon}, {Sato}, {Shinozaki}, {Shipman}, {Sugita}, {Suzuki}, {van der Tak}, {Torres Redondo}, {Wada}, {Wang}, {Wafelbakker}, {van Weers}, {Withington}, {Vandenbussche}, {Yamada}, \& {Yamamura}}]{roelfsema2018}
{Roelfsema}, P.~R., {Shibai}, H., {Armus}, L., {et~al.} 2018, \pasa, 35, e030

\bibitem[{{Romero-Mirza} {et~al.}(2024){Romero-Mirza}, {Banzatti}, {{\"O}berg}, {Pontoppidan}, {Salyk}, {Najita}, {Blake}, {Krijt}, {Arulanantham}, {Pinilla}, {Long}, {Rosotti}, {Andrews}, {Wilner}, {Calahan}, \& {JDISCS Collaboration}}]{romero-mirza2024_sample}
{Romero-Mirza}, C.~E., {Banzatti}, A., {{\"O}berg}, K.~I., {et~al.} 2024, \apj, 975, 78

\bibitem[{Rothman {et~al.}(2013)Rothman, Gordon, Babikov, Barbe, {Chris Benner}, Bernath, Birk, Bizzocchi, Boudon, Brown, Campargue, Chance, Cohen, Coudert, Devi, Drouin, Fayt, Flaud, Gamache, Harrison, Hartmann, Hill, Hodges, Jacquemart, Jolly, Lamouroux, {Le Roy}, Li, Long, Lyulin, Mackie, Massie, Mikhailenko, Müller, Naumenko, Nikitin, Orphal, Perevalov, Perrin, Polovtseva, Richard, Smith, Starikova, Sung, Tashkun, Tennyson, Toon, Tyuterev, \& Wagner}]{rothman2013}
Rothman, L., Gordon, I., Babikov, Y., {et~al.} 2013, Journal of Quantitative Spectroscopy and Radiative Transfer, 130, 4, hITRAN2012 special issue

\bibitem[{{Salyk} {et~al.}(2025){Salyk}, {Pontoppidan}, {Banzatti}, {Bergin}, {Arulanantham}, {Najita}, {Blake}, {Carr}, {Zhang}, \& {Xie}}]{salyk2025}
{Salyk}, C., {Pontoppidan}, K.~M., {Banzatti}, A., {et~al.} 2025, \aj, 169, 184

\bibitem[{{Salyk} {et~al.}(2011){Salyk}, {Pontoppidan}, {Blake}, {Najita}, \& {Carr}}]{salyk2011}
{Salyk}, C., {Pontoppidan}, K.~M., {Blake}, G.~A., {Najita}, J.~R., \& {Carr}, J.~S. 2011, \apj, 731, 130

\bibitem[{{Sellek} {et~al.}(2025){Sellek}, {Vlasblom}, \& {van Dishoeck}}]{sellek2024}
{Sellek}, A.~D., {Vlasblom}, M., \& {van Dishoeck}, E.~F. 2025, \aap, 694, A79

\bibitem[{{Semenov} {et~al.}(2006){Semenov}, {Wiebe}, \& {Henning}}]{semenov2006_turbulence}
{Semenov}, D., {Wiebe}, D., \& {Henning}, T. 2006, \apjl, 647, L57

\bibitem[{{Smith} {et~al.}(2025){Smith}, {Romero-Mirza}, {Banzatti}, {Rab}, {{\'A}brah{\'a}m}, {K{\'o}sp{\'a}l}, {Claes}, {Manara}, {{\"O}berg}, {Bouwman}, {de Miera}, \& {Green}}]{smith2025}
{Smith}, S.~A., {Romero-Mirza}, C.~E., {Banzatti}, A., {et~al.} 2025, \apjl, 984, L51

\bibitem[{Stacey {et~al.}(2024)Stacey, Anderson, Arendt, Banzatti, Baselmans, Bergin, Bergner, Bjoraker, Chen, Cleeves, Cothard, Dabironezare, Ferrari, Greenhouse, Jellema, Kovacs, Kutyrev, MacGregor, Melnick, Milam, Nikola, Pontoppidan, Vazquez, Rostem, Stahl, Thelen, Tolls, Trapman, Wollack, \& Zhang}]{poemm_spie}
Stacey, G.~J., Anderson, D., Arendt, R., {et~al.} 2024, in Millimeter, Submillimeter, and Far-Infrared Detectors and Instrumentation for Astronomy XII, ed. J.~Zmuidzinas \& J.-R. Gao, Vol. PC13102, International Society for Optics and Photonics (SPIE), PC131021G

\bibitem[{{Tabone} {et~al.}(2023){Tabone}, {Bettoni}, {van Dishoeck}, {Arabhavi}, {Grant}, {Gasman}, {Henning}, {Kamp}, {G{\"u}del}, {Lagage}, {Ray}, {Vandenbussche}, {Abergel}, {Absil}, {Argyriou}, {Barrado}, {Boccaletti}, {Bouwman}, {Garatti}, {Geers}, {Glauser}, {Justannont}, {Lahuis}, {Mueller}, {Nehm{\'e}}, {Olofsson}, {Pantin}, {Scheithauer}, {Waelkens}, {Waters}, {Black}, {Christiaens}, {Guadarrama}, {Morales-Calder{\'o}n}, {Jang}, {Kanwar}, {Pawellek}, {Perotti}, {Perrin}, {Rodgers-Lee}, {Samland}, {Schreiber}, {Schwarz}, {Colina}, {{\"O}stlin}, \& {Wright}}]{tabone2023}
{Tabone}, B., {Bettoni}, G., {van Dishoeck}, E.~F., {et~al.} 2023, Nature Astronomy

\bibitem[{{Temmink} {et~al.}(2025){Temmink}, {Sellek}, {Gasman}, {van Dishoeck}, {Vlasblom}, {Pranger}, {G{\"u}del}, {Henning}, {Lagage}, {Caratti o Garatti}, {Kamp}, {Olofsson}, {Arabhavi}, {Grant}, {Kaeufer}, {Kurtovic}, {Perotti}, {Samland}, {Schwarz}, \& {Tabone}}]{temmink2025}
{Temmink}, M., {Sellek}, A.~D., {Gasman}, D., {et~al.} 2025, \aap, 699, A134

\bibitem[{{Temmink} {et~al.}(2024){Temmink}, {van Dishoeck}, {Gasman}, {Grant}, {Tabone}, {G{\"u}del}, {Henning}, {Barrado}, {Caratti o Garatti}, {Glauser}, {Kamp}, {Arabhavi}, {Jang}, {Kurtovic}, {Perotti}, {Schwarz}, \& {Vlasblom}}]{temmink2024b}
{Temmink}, M., {van Dishoeck}, E.~F., {Gasman}, D., {et~al.} 2024, \aap, 689, A330

\bibitem[{{Tennyson} {et~al.}(2001){Tennyson}, {Zobov}, {Williamson}, {Polyansky}, \& {Bernath}}]{tennyson2001}
{Tennyson}, J., {Zobov}, N.~F., {Williamson}, R., {Polyansky}, O.~L., \& {Bernath}, P.~F. 2001, Journal of Physical and Chemical Reference Data, 30, 735

\bibitem[{{van Dishoeck} {et~al.}(2023){van Dishoeck}, {Grant}, {Tabone}, {van Gelder}, {Francis}, {Tychoniec}, {Bettoni}, {Arabhavi}, {Gasman}, {Nazari}, {Vlasblom}, {Kavanagh}, {Christiaens}, {Klaassen}, {Beuther}, {Henning}, \& {Kamp}}]{vandishoeck2023}
{van Dishoeck}, E.~F., {Grant}, S., {Tabone}, B., {et~al.} 2023, Faraday Discussions, 245, 52

\bibitem[{{van Dishoeck} {et~al.}(2013){van Dishoeck}, {Herbst}, \& {Neufeld}}]{vandishoeck2013}
{van Dishoeck}, E.~F., {Herbst}, E., \& {Neufeld}, D.~A. 2013, Chemical Reviews, 113, 9043

\bibitem[{{van Dishoeck} {et~al.}(2006){van Dishoeck}, {Jonkheid}, \& {van Hemert}}]{vandishoeck2006}
{van Dishoeck}, E.~F., {Jonkheid}, B., \& {van Hemert}, M.~C. 2006, Faraday Discussions, 133, 231

\bibitem[{{van Dishoeck} {et~al.}(2021){van Dishoeck}, {Kristensen}, {Mottram}, {Benz}, {Bergin}, {Caselli}, {Herpin}, {Hogerheijde}, {Johnstone}, {Liseau}, {Nisini}, {Tafalla}, {van der Tak}, {Wyrowski}, {Baudry}, {Benedettini}, {Bjerkeli}, {Blake}, {Braine}, {Bruderer}, {Cabrit}, {Cernicharo}, {Choi}, {Coutens}, {de Graauw}, {Dominik}, {Fedele}, {Fich}, {Fuente}, {Furuya}, {Goicoechea}, {Harsono}, {Helmich}, {Herczeg}, {Jacq}, {Karska}, {Kaufman}, {Keto}, {Lamberts}, {Larsson}, {Leurini}, {Lis}, {Melnick}, {Neufeld}, {Pagani}, {Persson}, {Shipman}, {Taquet}, {van Kempen}, {Walsh}, {Wampfler}, {Y{\i}ld{\i}z}, \& {WISH Team}}]{vandishoeck2021}
{van Dishoeck}, E.~F., {Kristensen}, L.~E., {Mottram}, J.~C., {et~al.} 2021, \aap, 648, A24

\bibitem[{{Villenave} {et~al.}(2025){Villenave}, {Rosotti}, {Lambrechts}, {Ziampras}, {Pinte}, {M{\'e}nard}, {Stapelfeldt}, {Duch{\^e}ne}, {Baylock}, \& {Doi}}]{villenave2025}
{Villenave}, M., {Rosotti}, G.~P., {Lambrechts}, M., {et~al.} 2025, \aap, 697, A64

\bibitem[{{Visser} {et~al.}(2018){Visser}, {Bruderer}, {Cazzoletti}, {Facchini}, {Heays}, \& {van Dishoeck}}]{visser2018}
{Visser}, R., {Bruderer}, S., {Cazzoletti}, P., {et~al.} 2018, \aap, 615, A75

\bibitem[{{Vlasblom} {et~al.}(2025){Vlasblom}, {Temmink}, {Grant}, {Kurtovic}, {Sellek}, {van Dishoeck}, {G{\"u}del}, {Henning}, {Lagage}, {Barrado}, {Caratti o Garatti}, {Glauser}, {Kamp}, {Lahuis}, {Olofsson}, {Arabhavi}, {Christiaens}, {Gasman}, {Jang}, {Morales-Calder{\'o}n}, {Perotti}, {Schwarz}, \& {Tabone}}]{vlasblom2025}
{Vlasblom}, M., {Temmink}, M., {Grant}, S.~L., {et~al.} 2025, \aap, 693, A278

\bibitem[{{Vlasblom} {et~al.}(2024){Vlasblom}, {van Dishoeck}, {Tabone}, \& {Bruderer}}]{vlasblom2024}
{Vlasblom}, M., {van Dishoeck}, E.~F., {Tabone}, B., \& {Bruderer}, S. 2024, \aap, 682, A91

\bibitem[{{Walsh} {et~al.}(2015){Walsh}, {Nomura}, \& {van Dishoeck}}]{walsh2015}
{Walsh}, C., {Nomura}, H., \& {van Dishoeck}, E. 2015, \aap, 582, A88

\bibitem[{{Wilson} \& {Rood}(1994)}]{wilson1994}
{Wilson}, T.~L. \& {Rood}, R. 1994, \araa, 32, 191

\bibitem[{{Woitke} {et~al.}(2022){Woitke}, {Arabhavi}, {Kamp}, \& {Thi}}]{woitke2022}
{Woitke}, P., {Arabhavi}, A.~M., {Kamp}, I., \& {Thi}, W.~F. 2022, \aap, 668, A164

\bibitem[{{Woitke} {et~al.}(2009){Woitke}, {Kamp}, \& {Thi}}]{woitke2009}
{Woitke}, P., {Kamp}, I., \& {Thi}, W.~F. 2009, \aap, 501, 383

\bibitem[{{Woitke} {et~al.}(2018){Woitke}, {Min}, {Thi}, {Roberts}, {Carmona}, {Kamp}, {M{\'e}nard}, \& {Pinte}}]{woitke2018}
{Woitke}, P., {Min}, M., {Thi}, W.~F., {et~al.} 2018, \aap, 618, A57

\bibitem[{{Woitke} {et~al.}(2024){Woitke}, {Thi}, {Arabhavi}, {Kamp}, {K{\'o}sp{\'a}l}, \& {{\'A}brah{\'a}m}}]{woitke2024}
{Woitke}, P., {Thi}, W.~F., {Arabhavi}, A.~M., {et~al.} 2024, \aap, 683, A219

\bibitem[{{Woodall} {et~al.}(2007){Woodall}, {Ag{\'u}ndez}, {Markwick-Kemper}, \& {Millar}}]{woodall2007}
{Woodall}, J., {Ag{\'u}ndez}, M., {Markwick-Kemper}, A.~J., \& {Millar}, T.~J. 2007, \aap, 466, 1197

\bibitem[{{Xie} {et~al.}(1995){Xie}, {Allen}, \& {Langer}}]{xie1995_turbulence}
{Xie}, T., {Allen}, M., \& {Langer}, W.~D. 1995, \apj, 440, 674

\bibitem[{{Xu} {et~al.}(2019){Xu}, {Bai}, {{\"O}berg}, \& {Zhang}}]{xu2019}
{Xu}, R., {Bai}, X.-N., {{\"O}berg}, K., \& {Zhang}, H. 2019, \apj, 872, 107

\bibitem[{{Zhang} {et~al.}(2021){Zhang}, {Booth}, {Law}, {Bosman}, {Schwarz}, {Bergin}, {{\"O}berg}, {Andrews}, {Guzm{\'a}n}, {Walsh}, {Qi}, {van't Hoff}, {Long}, {Wilner}, {Huang}, {Czekala}, {Ilee}, {Cataldi}, {Bergner}, {Aikawa}, {Teague}, {Bae}, {Loomis}, {Calahan}, {Alarc{\'o}n}, {M{\'e}nard}, {Le Gal}, {Sierra}, {Yamato}, {Nomura}, {Tsukagoshi}, {P{\'e}rez}, {Trapman}, {Liu}, \& {Furuya}}]{zhang2021}
{Zhang}, K., {Booth}, A.~S., {Law}, C.~J., {et~al.} 2021, \apjs, 257, 5

\bibitem[{{Zhang} {et~al.}(2013){Zhang}, {Pontoppidan}, {Salyk}, \& {Blake}}]{zhang2013}
{Zhang}, K., {Pontoppidan}, K.~M., {Salyk}, C., \& {Blake}, G.~A. 2013, \apj, 766, 82

\end{thebibliography}

\begin{appendix}
\section{Additional model explorations}\label{app}

\subsection{Retrievals on \ce{CO2} emission}\label{app:CO2}

Since \ce{CO2} emission is seen in many disks observed with JWST/MIRI \citep[e.g.,][]{grant2023, vlasblom2025, salyk2025}, we also investigate how the temperature and column density retrieved from LTE slab fits on synthetic \ce{CO2} spectra compare to our models. Just as for \ce{H2O}, we perform our retrievals both on the model using the full chemistry, and on a model with a parameterized abundance. The abundance maps for these models are shown in Fig. \ref{fig:abu_fl9_CO2}. The 70\% emitting regions of the \ce{^12CO2} and \ce{^13CO2} $\nu = 1-0$ $Q$(20) lines (green and purple, respectively) are also indicated. These species trace further out into the disk than the \ce{H2O} lines (11$_{3,9}$ -- 10$_{0,10}$ line with $E_{\rm up}$ = 2438 K is indicated in pink), and the \ce{^13CO2} traces slightly deeper down into the disk than the \ce{^12CO2}, due to it being more optically thin. The reason that \ce{CO2} traces further out than \ce{H2O} is most likely due to \ce{CO2} being more efficiently formed at lower temperatures, where the \ce{OH + CO -> CO2 + H} reaction is more efficient than the \ce{OH + H2 -> H2O + H} reaction \citep{charnley1997, vandishoeck2013, walsh2015}. \\
\newline
Additionally, the version of DALI used in this work includes the effects of UV shielding by \ce{H2O} \citep{bethell2009} in its calculation of the chemistry and thermal balance, as implemented in \citet{bosman2022a}. This process is most efficient in the innermost regions of the disk, where the \ce{H2O} column density can build up substantially due to efficient gas-phase formation. As a result, the formation of \ce{CO2} is strongly quenched in these regions, as the self-shielding of \ce{H2O} prevents the production of OH via \ce{H2O} photodissociation, thus preventing \ce{CO2} production. Additionally, the photodissociation rate of \ce{CO2} is higher in these innermost regions as it cannot benefit from self-shielding at these column densities. This creates a gap in the \ce{CO2} abundance within 0.4 au, as first demonstrated in \citet{bosman2022b}, and also seen in our model (left panel of Fig. \ref{fig:abu_fl9}). \\
\newline
For consistency, this effect is recreated in the parameterized abundance model and as such, we use a more complex abundance structure to better match the model with the full chemistry. We define the high-abundance reservoir as the region where 200 K > $T_{\rm dust}$ > 70 K and $A_V$ > 5 mag, where we set the fractional abundance to $n$(\ce{CO2})/$n_{\rm H}$ = $5\times10^{-5}$. We also define a lower-abundance surface layer as the region where 1 < $A_V$ < 5 and $T_{\rm dust}$ > 70 K, where we set the fractional abundance to $n$(\ce{CO2})/$n_{\rm H}$ = $10^{-7}$. The fractional \ce{CO2} abundance in the rest of the disk is set to $10^{-10}$. \\
\newline
Since \ce{^12CO2} and \ce{^13CO2} each only have one main emission feature in the MIRI wavelength range, we fit these features with a single LTE slab model between 13.5 and 17.5 $\mu$m, retrieving a single best-fit value for $T$, $N$ and $R_{\rm eq}$. The best-fit model was obtained from a grid of models using the $\chi^2$ method described in Sect. \ref{sec:methods}. In this grid, $T$ was varied linearly between 100 and 1000 K in steps of $\Delta T = 22.5 K$ and $N$ was varied in log-space between 10$^{14}$ and 10$^{23}$ cm$^{-2}$ with steps of $\Delta \log N = 0.225$. For \ce{^13CO2}, $T$ was varied between 50 and 500 K in steps of $\Delta T = 11.25$ K. \\
\newline
The results of these fits are shown in Fig. \ref{fig:CBF_Tgas_fl9_CO2}, and just like for \ce{H2O}, the agreement between the retrieved temperature and the temperature within the emitting layer of the DALI model is quite good. We find that the \ce{^13CO2} emission traces a slightly lower temperature than the \ce{^12CO2}, likely due to it tracing deeper down into the disk. The retrieved column densities are presented in Fig. \ref{fig:N_rad_fl9_CO2}, which trace close to the $\tau=1$ surface, just as was found for \ce{H2O}.

\begin{table}[]
    \centering
    \caption{Lines analyzed in this work.}
    \begin{tabular}{c c c c c}
        \hline \hline
         Molecule & Transition & $\lambda$ ($\mu$m) & $E_{\rm up}$ (K) & $A_{\rm ul}$ (s$^{-1}$) \\
         \hline 
        \ce{H2O}        & $17_{7,10} - 16_{4,13}$ & 10.11 & 6371 & 1.56 \\
                        & $11_{3,9} - 10_{0,10}$ & 17.22 & 2438 & 0.99 \\
                        & $16_{8,9} - 15_{7,8}$ & 17.32 & 6052\tablefootmark{a} & 41.5 \\
                        & $13_{4,9} - 12_{3,10}$ & 17.50 & 3645\tablefootmark{a} & 4.94 \\
                        & $8_{3,6} - 7_{0,7}$ & 23.81 & 1447\tablefootmark{a} & 0.61 \\
                        & $8_{4,5} - 7_{1,6}$ & 23.89 & 1615\tablefootmark{a} & 0.99 \\
        &&&&\\
        \ce{^12CO2}    & $\nu_2 = 1-0$ Q(20) & 14.97 & 1196 & 1.54\\ 
        &&&&\\
        \ce{^13CO2}     & $\nu_2 = 1-0$ Q(20) & 15.41 & 1169 & 1.38 \\
         \hline
    \end{tabular}
    \tablefoot{
    \tablefoottext{a}{Temperature diagnostic line identified in \citet{banzatti2025}}
    }
    \label{tab:Lines}
\end{table}

\subsection{LTE vs. non-LTE excitation} \label{app:LTE}

As mentioned in Sect. \ref{sec:methods}, the excitation in DALI can be calculated either in LTE or non-LTE, if collisional data are available for the species in question. In the main body of this work, the excitation for \ce{H2O} is calculated in non-LTE, from which the synthetic spectra are then generated. However, all subsequent retrievals on these synthetic spectra assume LTE excitation. To understand the implications of this, we calculate the excitation for our two fiducial models (full chemistry and parameterized abundances) in LTE as well, generate new synthetic spectra and run our LTE retrievals again. The results are presented in Fig. \ref{fig:CBF_Tgas_fl9_LTE}.\\
\newline
Fig. \ref{fig:CBF_Tgas_fl9_LTE} shows the retrieved temperatures of our slab fits in three wavelength ranges (10-14, 13.5-17.5, and 21-24 $\mu$m) for both our LTE (opaque crosses) and non-LTE (faded crosses) excitation models. The temperature retrieved from our LTE model is slightly higher than the temperature retrieved from our non-LTE (fiducial) model, both in the full chemistry and parameterized case. We note that this effect is also seen for the three-component MCMC fit, though this is not shown on Fig. \ref{fig:CBF_Tgas_fl9_LTE}. This clearly indicates that, if non-LTE effects are accounted for, the \ce{H2O} emission seen with JWST is slightly sub-thermally excited, as was also found in previous work \citep{meijerink2009}. As such, the retrieved temperature from an observation can in reality be slightly too low (or the emitting radius should be slightly larger), though the discrepancy is relatively small at only $\sim$100 K.

\subsection{Increased gas-to-dust ratio}\label{app:fl999}

It is predicted that the gas-to-dust ratio in the IR emitting layer is elevated above the canonical ISM value of 100 \citep{meijerink2009}. In DALI, this can be parameterized either by the gas-to-dust mass ratio (which is kept at 100 for all models in this work), or by the parameter $f_\ell$, which sets the mass fraction of the large grain population. Since the large grains have a reduced scale height with respect to the small grains (which follow the vertical extent of the gas), a larger value for $f_\ell$ increases the gas-to-dust ratio in the upper layers of the disk. In the models presented in the main body of this work, $f_\ell$ is set to 0.9, which yields a gas-to-dust ratio of $\sim10^3$ in the IR emitting layers. We run both fiducial models (full chemistry and parameterized abundances) with $f_\ell=0.99$ and 0.999, yielding even higher gas-to-dust ratios of $10^4$ and $10^5$ in the upper layers of the disk. \\
\newline
Fig. \ref{fig:CBF_Tgas_fl999} presents the retrieved temperatures for the $f_\ell=0.999$ models. They largely follow the same conclusions as the fiducial models: the single-temperature fits mainly trace the warm $\sim$500 K component, whereas the three-component MCMC allows for the full temperature gradient present in the emitting region to be captured better. We note that the full chemistry model with $f_\ell=0.999$ is quite warm, and thus the contribution from the coldest (<300 K) component is very weak, meaning that this component of the MCMC is poorly constrained. \\
\newline
Fig. \ref{fig:N_rad_fl999} also presents the retrieved column densities. Whereas these mainly traced the $\tau=1$ surface in the fiducial models, in this model we see that this is not the case. Since the large dust grains are so strongly settled to the midplane in this model, the $\tau=1$ surface lays very deep into the disk, and thus the total observable column of gas is large. As such, the \ce{H2O} gas is able to become optically thick before the dust does, and the retrieved column densities no longer trace the dust $\tau=1$ surface, but rather the gas $\tau=1$ surface. This is similar to what was seen for HCN by \citet{bruderer2015}.

\subsection{Influence of the gas temperature}\label{app:tgas}

In the main body of this work, we keep the temperature structure of the disk constant between our full chemistry and parameterized models, which allows us to isolate the effects of changes to the abundance structure. Here, we explore how the gas temperature, as well as the strength of the radiation field, influences the \ce{H2O} abundance in the IR emitting layer. \\
\newline
For this analysis, we construct a very simple chemical network following \citet{kristensen2017}. We assume that all \ce{H2O} is formed through gas-phase reactions with \ce{H2}. This is governed by two reactions: \ce{O + H2 -> OH + H} and \ce{OH + H2 -> H2O + H}. We also assume that \ce{H2O} is only destroyed through photodissociation, where it is rapidly dissociated all the way back to atomic O (this can happen either in a single step or in two steps, where OH dissociates into O + H right after the \ce{H2O} photodissociation). The rate of photodissociation is given by $R_{\rm pd} = G_0 k_{\rm pd} n_{\ce{H2O}}$, where $G_0$ is the radiation field strength in units of the interstellar radiation field, $k_{\rm pd} = 8.0 \times 10^{-10}$ s$^{-1}$ \citep{vandishoeck2006} and $n_{\ce{H2O}}$ is the \ce{H2O} abundance. Thus, assuming equilibrium, we find that 
\begin{align}
    \frac{ {\rm d}n_{\ce{H2O}} }{ {\rm d}t } = k_{\ce{H2O}}n_{\ce{H2}}n_{\ce{OH}} - G_0k_{\rm pd}n_{\ce{H2O}} = 0.
\end{align}
Regarding OH, we assume this to be rapidly produced from \ce{O + H2 -> OH + H} and destroyed through \ce{OH + H2 -> H2O + H} (we do not consider OH photodissociation), and thus the molecule will not have a substantial abundance with respect to O and \ce{H2O}. Thus, in equilibrium,
\begin{align}
    \frac{ {\rm d}n_{\ce{OH}} }{ {\rm d}t } = k_{\ce{OH}}n_{\ce{H2}}n_{\ce{O}} - k_{\ce{H2O}}n_{\ce{H2}}n_{\ce{OH}} = 0.
\end{align}
From these two equations, we find that
\begin{align}
    n_{\ce{H2O}} = \frac{ k_{\ce{OH}} }{ k_{\rm pd}} \frac{ n_{\ce{H2}}n_{\ce{O}} }{ G_0 }.
\end{align}
Finally, we assume that $n_{\rm O} = 2.88\times10^{-4}\,n_{\ce{H2}}$, following the assumed O/H abundance in our models, and we find that $k_{\rm OH}(T) = 3.13\times10^{-13} (T/300\,{\rm K})^{2.7}\exp(-3150\,{\rm K}/T)$ \citep[UMIST22;][]{millar2024}.\\
\newline
In Fig. \ref{fig:H2_G0_Tgas}, we present the \ce{H2} density, radiation field strength $G_0$, and gas temperature in our fiducial chemistry model. The solid contour indicates our parameterized \ce{H2O} abundance structure, to illustrate what the relevant conditions are in the different parts in the IR emitting layer. For this analysis, we divide the \ce{H2O} reservoir into two parts, just as was done in Sect. \ref{subsec:res_H2O}: an inner, warm reservoir that is located within 1 au, and an outer, colder reservoir outside 1 au. We only consider the upper, emitting layers above Z/R = 0.15. Fig. \ref{fig:H2_G0_Tgas} demonstrates that the inner reservoir has a relatively high \ce{H2} density ($\sim10^{10}-10^{11}$ cm$^{-3}$; see \citealt{vandishoeck2021} for a lower density case), a radiation field between $10^4-10^6\,G_0$ and a temperature $\gtrsim 300$ K. The outer reservoir, on the other hand, has a slightly lower \ce{H2} density, a slightly weaker radiation field (between $10^3-10^5\,G_0$) and a lower temperature $\lesssim400$ K. \\
\newline
Fig. \ref{fig:nH2O} presents the expected \ce{H2O} abundance as a function of temperature and $G_0$ from our calculation, where the red and blue boxes represent the estimated conditions of the inner and outer \ce{H2O} reservoir, as specified above. This figure clearly demonstrates that building up abundant \ce{H2O} ($n$(\ce{H2O})/$n_{\rm H} \sim 10^{-4}$) is easily attainable in the inner, warmest regions of the disk (especially since the \ce{H2} density is also higher), but very hard to attain in the outer reservoir, as the temperature is clearly too low. \\
\newline
Thus, Fig. \ref{fig:abu_Tgas} presents a scenario in which the disk gas temperature has been strongly increased throughout the entire disk, by a factor 2. We do not increase the dust temperature, thus the snow surface remains in the same place. The increase in gas temperature allows \ce{H2O} to form abundantly out to much larger radii and the abundance structure closely resembles our parameterized model (see Fig. \ref{fig:abu_fl9}). Still, the spectrum in the third panel of Fig. \ref{fig:abu_Tgas} demonstrates that the enhancement in the cold \ce{H2O} lines is much less than what was attained by our parameterized model. After all, while the cold \ce{H2O} is slightly enhanced due to an increase in its emitting area, the warm and hot \ce{H2O} have undergone a similar increase in their emitting area, and thus the relative strength of the cold lines has not changed much. \\
\newline
As such, a spectrum dominated by cold \ce{H2O} seems to require a high \ce{H2O} abundance in the layer above the snow surface, along with a low temperature $\lesssim$300 K. Such a reservoir is not present in this model, as the lowest-temperature gas-phase \ce{H2O} of the fiducial model at $\sim$150 K is now twice as hot. Since Fig. \ref{fig:abu_Tgas} has demonstrated that gas-phase formation alone at temperatures below 300 K does not allow for abundant \ce{H2O} to be built up, this leads to one of two conclusions: 1) the \ce{H2O} must remain abundant after the disk cools down from a heating event (such as an accretion outburst, see also Sect. \ref{subsec:accretion}), which can potentially be aided by UV shielding, or 2) the \ce{H2O} must become abundant through some other mechanism than gas-phase formation, such as the sublimation of icy pebbles followed by vertical mixing. The latter may also require some form of continuous replenishment due to the short photodissociation timescales in the upper layers. We note that the effects of additional ice sublimation are not included in this model, as the dust temperature was not enhanced along with the gas temperature. Regardless, the \ce{H2O} ice reservoir is located below Z/R = 0.2, below the cold \ce{H2O} emitting region and would thus not contribute to its emission unless it is mixed upwards.

\begin{table*}[t]
    \caption{Summary of model grids and parameters.} 
    \centering
    \begin{tabular}{l c l}
        \hline
        \hline
         Parameter          & Symbol            & Value  \\
         \hline
         & Disk structure - all models & \\
        \hline
        Sublimation radius  & $R_{\rm subl}$    & 0.08 au    \\  
        Outer radius        & $R_{\rm out}$     & 100 au \\
        Characteristic radius   & $R_{\rm c}$         & 46 au \\
        Characteristic gas surface density   & $\Sigma_{\rm c}$    & 21.32 g cm$^{-2}$ \\
        Power-law index     & $\gamma$          & 0.9 \\
        Characteristic scale height  & $h_{\rm c}$             & 0.08 \\
        Flaring index       & $\psi$            & 0.11 \\
        Large dust mass fraction & $f_\ell$             & [\textbf{0.9}, 0.99, 0.999] \\ 
        Large dust settling parameter              & $\chi$            & 0.2 \\
        \hline
        & Stellar parameters - chem. \& param. models & \\
        \hline
        Input stellar spectrum & - & AS 209\tablefootmark{a} \\
        Stellar luminosity          & $L_*$             & 1.4 $L_\odot$\\
        Effective temperature     & $T_{\rm eff}$     & 4300 K \\
        FUV luminosity & $\log L_{\rm FUV}/L_\odot$     & -2.12 \\
        \hline
        & Stellar parameters - FUV models & \\
        \hline
        Input stellar spectrum & - & 4000 K blackbody + 10$\,$000 K excess\tablefootmark{b} \\
        Stellar luminosity          & $L_*$             & 0.8 $L_\odot$\\
        Accretion luminosity          & $L_{\rm acc}$           & [0, 0.002, 0.02, 0.2] $L_\odot$\\
        Mass accretion rate          & $\dot{M}_{\rm acc}$     & [0, 10$^{-10}$, 10$^{-9}$, 10$^{-8}$] $M_\odot$ yr$^{-1}$\\
        FUV luminosity & $\log L_{\rm FUV}/L_\odot$     & [-4.68, -3.75, -2.82, -1.82] \\
        \hline
        & Elemental abundances - chem. \& FUV models &\\
        \hline
        C abundance         & C/H          &$1.35\times10^{-4}$\\
        O abundance         & O/H          &$2.88\times10^{-4}$\\
        \hline
        & Parameterized abundances - param. models &\\
        \hline
        \ce{H2O} abundance - inner reservoir\tablefootmark{c} & $n$(\ce{H2O})/$n_{\rm H}$ & $10^{-4}$ \\
        \ce{H2O} abundance - outer reservoir\tablefootmark{c} & $n$(\ce{H2O})/$n_{\rm H}$ & [$10^{-6}$, $10^{-5}$, \textbf{10$^{-4}$}] \\
        \ce{CO2} abundance - upper reservoir\tablefootmark{d} & $n$(\ce{CO2})/$n_{\rm H}$ & $10^{-7}$ \\
        \ce{CO2} abundance - lower reservoir\tablefootmark{d} & $n$(\ce{CO2})/$n_{\rm H}$ & $5\times10^{-5}$ \\
    \end{tabular}
    \tablefoot{We divide the model parameters into categories and we indicate which models these parameters apply to. The disk structure is kept consistent between all models presented in this work. We highlight in boldface the fiducial $f_\ell$ value of 0.9 (models with the other values are presented in Sect. \ref{app:fl999} and Fig. \ref{fig:B25_plot}). All models except those presented in Sect. \ref{subsec:low_FUV} (the `FUV' models) use the AS 209 spectrum as input stellar spectrum. The FUV models instead use a blackbody with UV excess scaled to the mass accretion rate. \tablefoottext{a}{\citet{zhang2021}} \tablefoottext{b}{See \citet{kama2016, visser2018}} \tablefoottext{c}{See Sect. \ref{sec:methods}} \tablefoottext{d}{See Sect. \ref{app:CO2}}}
    \label{tab:params}
\end{table*}

\begin{figure*}
    \centering
    \includegraphics[width=\linewidth]{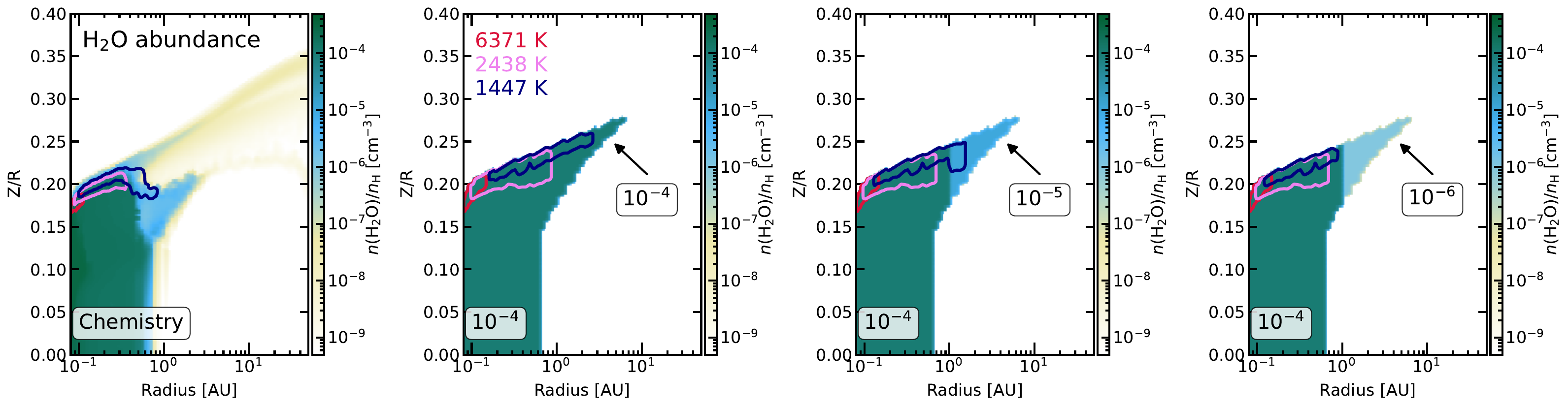}
    \caption{\ce{H2O} abundance maps of the fiducial model using the chemical network (first panel) and three models with parameterized abundances of $10^{-4}$, $10^{-5}$, and $10^{-6}$ beyond 1 au (second, third, and fourth panels). The $\tau=1$ surface of the dust continuum at 15 $\mu$m is indicated with a blue dashed line. Red, pink, and blue contours represent the 70\% emitting regions of the \ce{H2O} 17$_{7,10}$ -- 16$_{4,13}$ ($E_{\rm up} = 6371$ K), 11$_{3,9}$ -- 10$_{0,10}$ ($E_{\rm up} = 2438$ K), and 8$_{3,6}$ -- 7$_{0,7}$ ($E_{\rm up} = 1447$ K) lines, respectively. }
    \label{fig:Abu_H2O_depl}
\end{figure*}

\begin{figure*}
    \centering
    \includegraphics[width=0.7\linewidth]{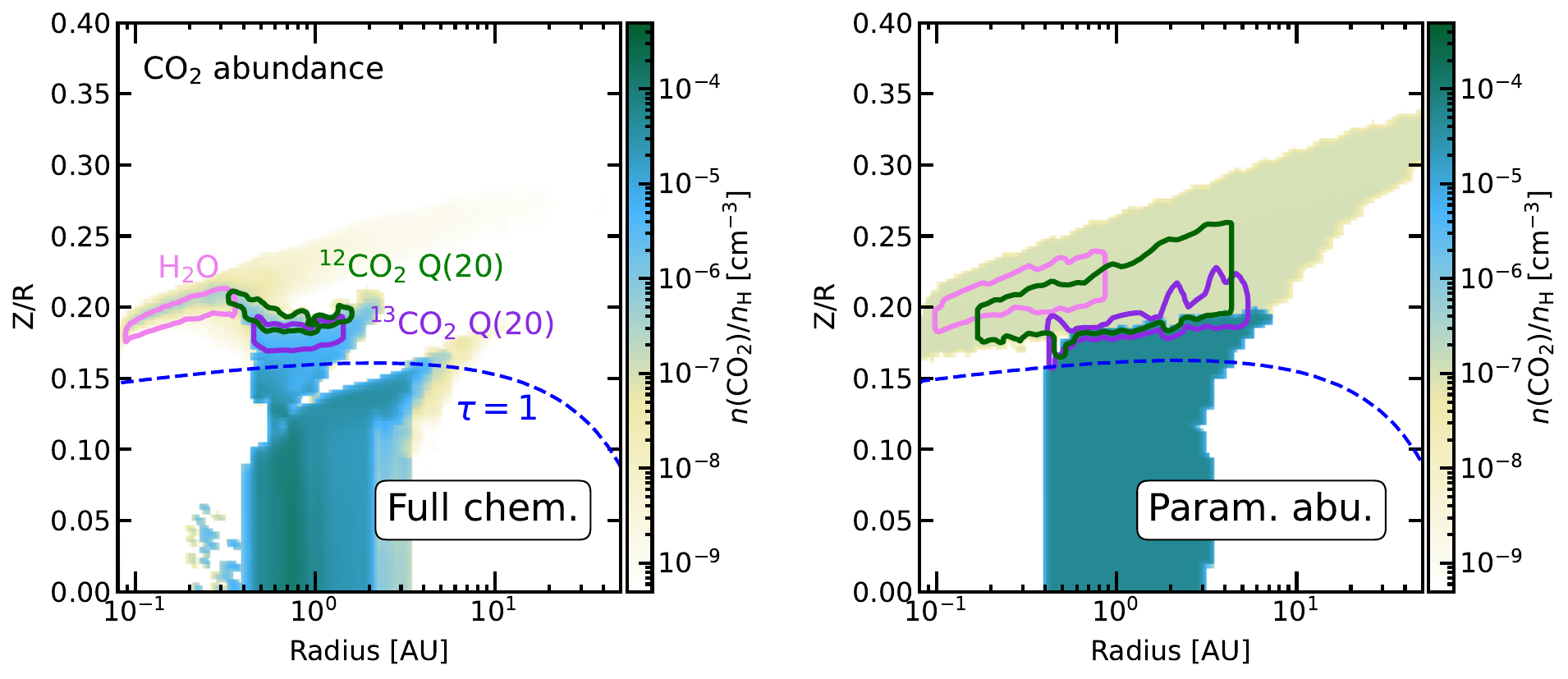}
    \caption{Abundance maps of \ce{CO2} for the fiducial models with $f_\ell=0.9$. The left panel depicts the model using the chemical network and the right panel depicts the model with parameterized abundances. The $\tau=1$ surface of the dust continuum at 15 $\mu$m is indicated with a blue dashed line in all panels. The green and purple contours represent the 70\% emitting regions of the \ce{^12CO2} and \ce{^13CO2} $\nu_2$ = 1 -- 0 $Q$(20) lines, respectively, and the pink contours represent the 70\% emitting region of the \ce{H2O} 11$_{3,9}$ -- 10$_{0,10}$ (2438 K) line. }
    \label{fig:abu_fl9_CO2}
\end{figure*}

\begin{figure*}
    \centering
    \includegraphics[width=0.65\linewidth]{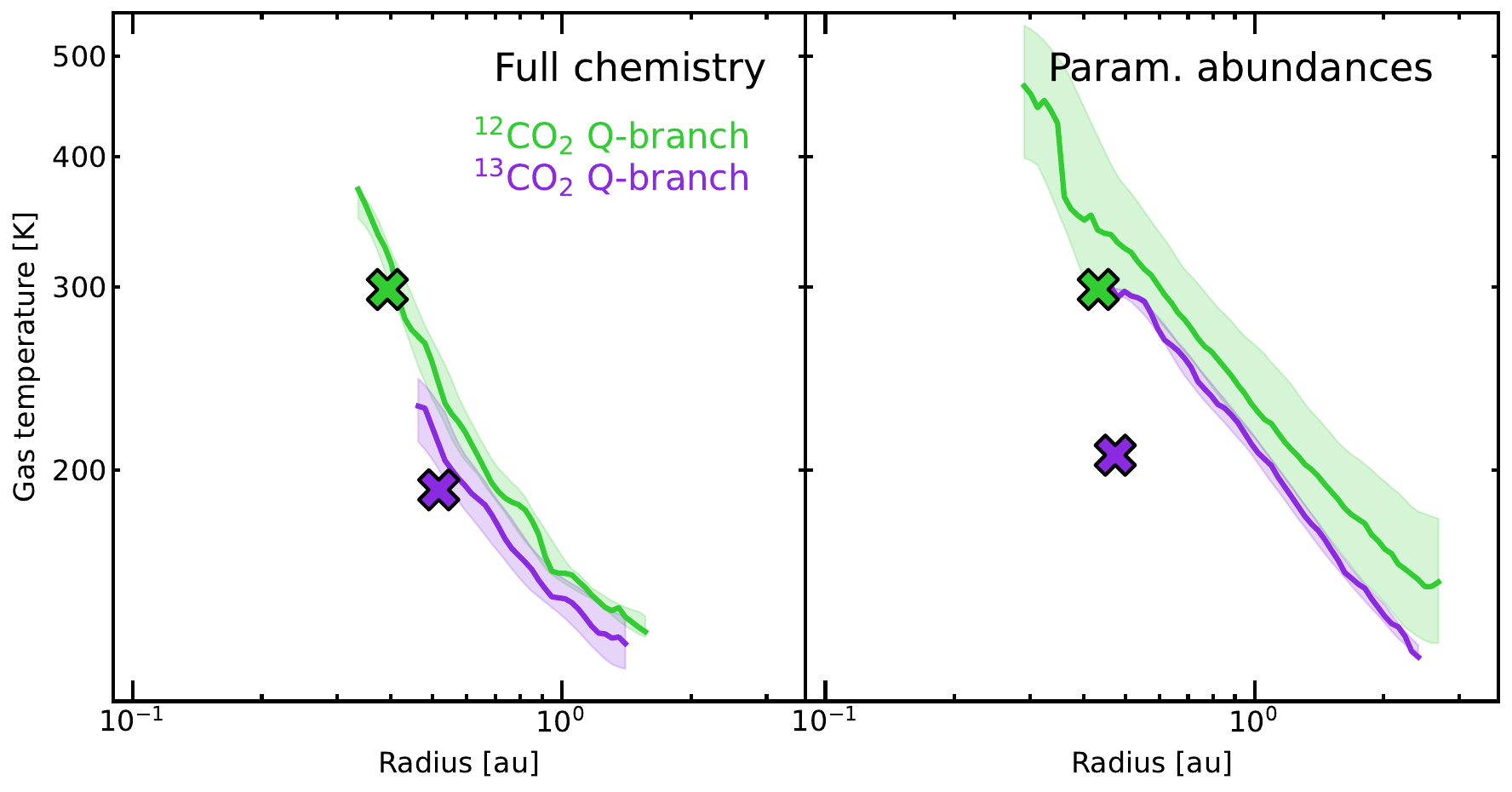}
    \caption{Temperature as a function of radius within the 70\% emitting region of the \ce{^12CO2} and \ce{^13CO2} $\nu_2$ = 1 -- 0 $Q$(20) lines (green and purple solid lines). The shaded regions represent the minimum and maximum temperature within the emitting region. The green and purple crosses represent the retrieved $T$ and $R_{\rm eq}$ for \ce{^12CO2} and \ce{^13CO2}. The left panel depicts the model using the chemical network and the right panel depicts the model with parameterized abundances. }
    \label{fig:CBF_Tgas_fl9_CO2}
\end{figure*}

\begin{figure*}
    \centering
    \includegraphics[width=0.8\linewidth]{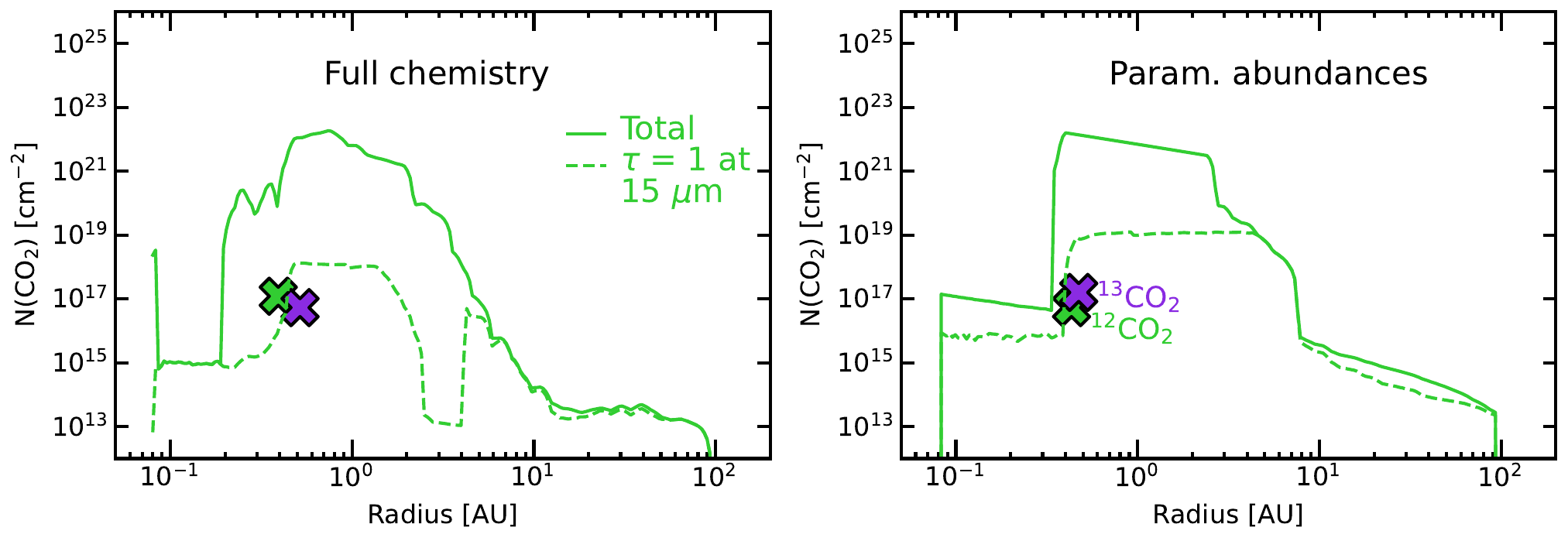}
    \caption{Vertically integrated \ce{CO2} column density as a function of radius. The solid lines show the total model column density and the dashed lines show the model column density integrated up to the dust $\tau=1$ surface at 15 $\mu$m. The green and purple crosses represent the retrieved $N$ and $R$ for \ce{^12CO2} and \ce{^13CO2}. The left panel depicts the model using the chemical network and the right panel depicts the model with parameterized abundances.}
    \label{fig:N_rad_fl9_CO2}
\end{figure*}

\begin{figure*}
    \centering
    \includegraphics[width=0.65\linewidth]{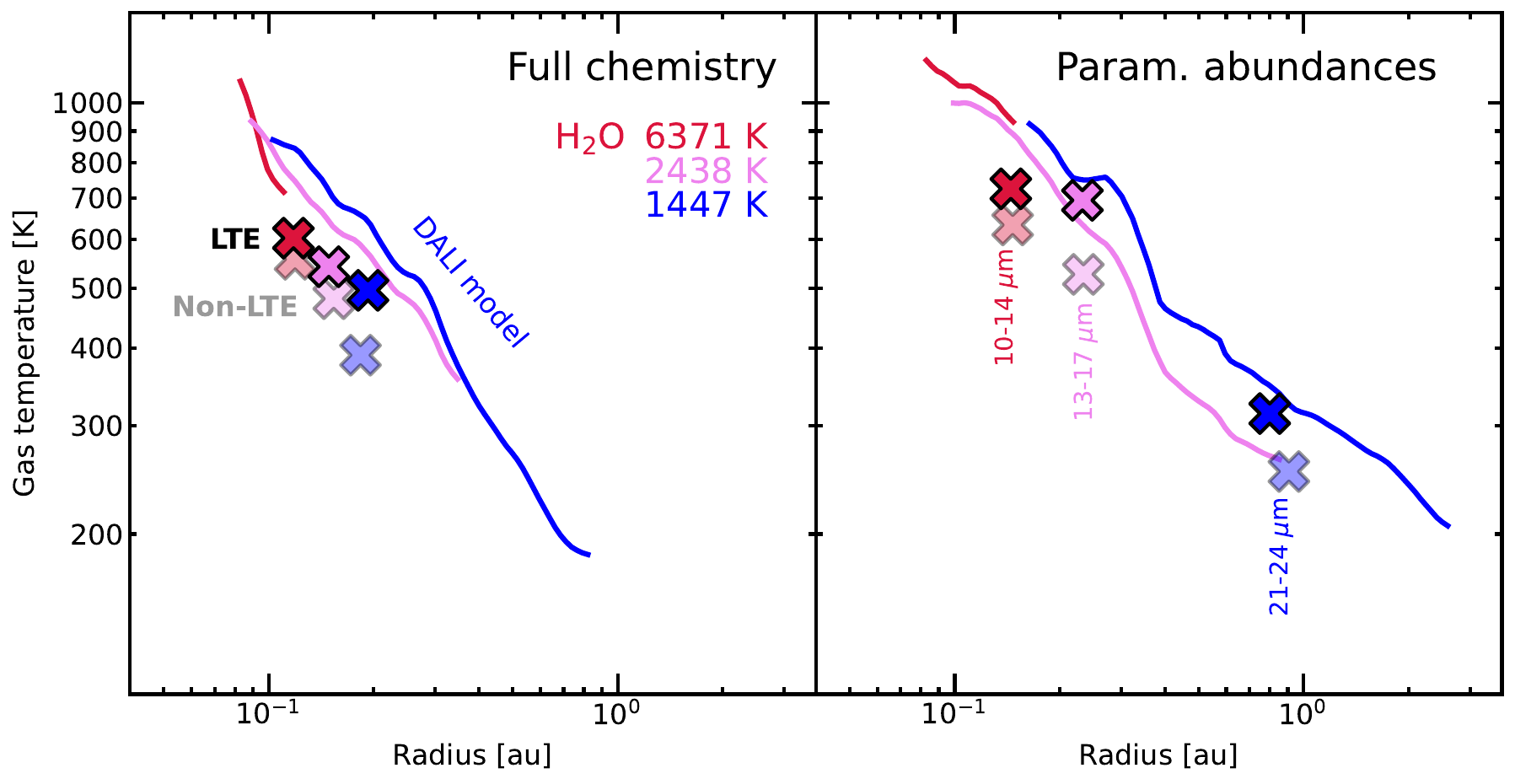}
    \caption{Temperature as a function of radius within the 70\% emitting region of the \ce{H2O} 17$_{7,10}$ -- 16$_{4,13}$ (6371 K), 11$_{3,9}$ -- 10$_{0,10}$ (2438 K), and 8$_{3,6}$ -- 7$_{0,7}$ (1447 K) lines (red, pink, and blue solid lines) for the model using the chemical network (left panel) and the model with parameterized abundances (right panel) with the excitation calculated in LTE. The red, pink, and blue crosses represent the retrieved $T$ and $R_{\rm eq}$ from the single-temperature slab fits in the 10-14, 13.5-17.5, and 21-24 $\mu$m region. The opaque crosses represent the fits done on the LTE-excitation models, and the faded crosses represent those done on the non-LTE excitation models (corresponding to the fits shown in Fig. \ref{fig:CBF_Tgas_fl9}).}
    \label{fig:CBF_Tgas_fl9_LTE}
\end{figure*}

\begin{figure*}
    \centering
    \includegraphics[width=0.65\linewidth]{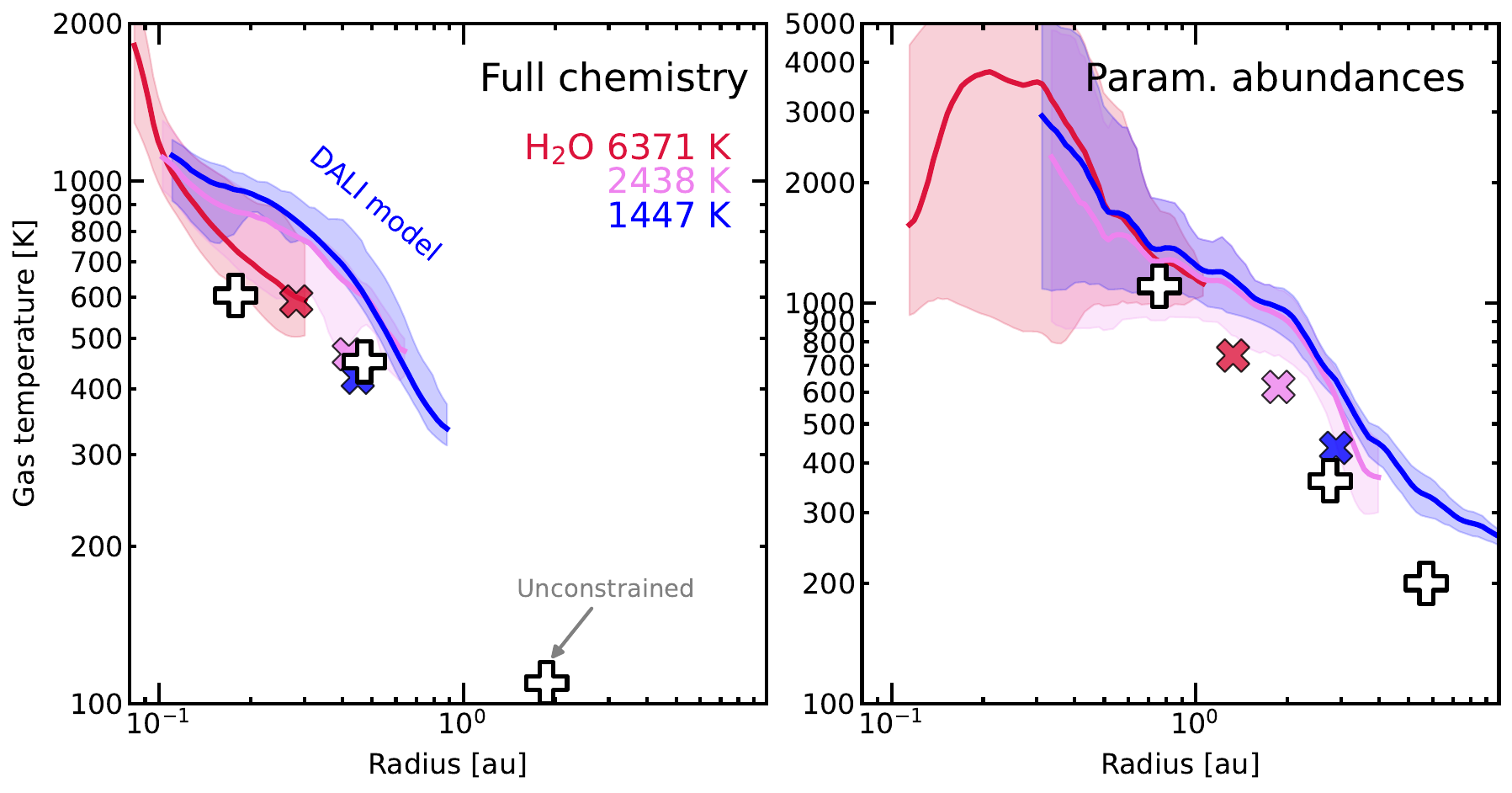}
    \caption{Temperature as a function of radius within the 70\% emitting region of the \ce{H2O} 17$_{7,10}$ -- 16$_{4,13}$ (6371 K), 11$_{3,9}$ -- 10$_{0,10}$ (2438 K), and 8$_{3,6}$ -- 7$_{0,7}$ (1447 K) lines (red, pink, and blue solid lines) for the models with $f_\ell=0.999$ with the full chemistry (left panel) and with parameterized abundances (right panel). The shaded regions represent the minimum and maximum temperature within the emitting region. The red, pink, and blue crosses represent the retrieved $T$ and $R_{\rm eq}$ from the single-temperature slab fits in the 10-14, 13.5-17.5, and 21-24 $\mu$m region. The white plus symbols represent the retrieved $T$ and $R_{\rm eq}$ values from the 3-temperature-component MCMC routine.}
    \label{fig:CBF_Tgas_fl999}
\end{figure*}

\begin{figure*}
    \centering
    \includegraphics[width=0.8\linewidth]{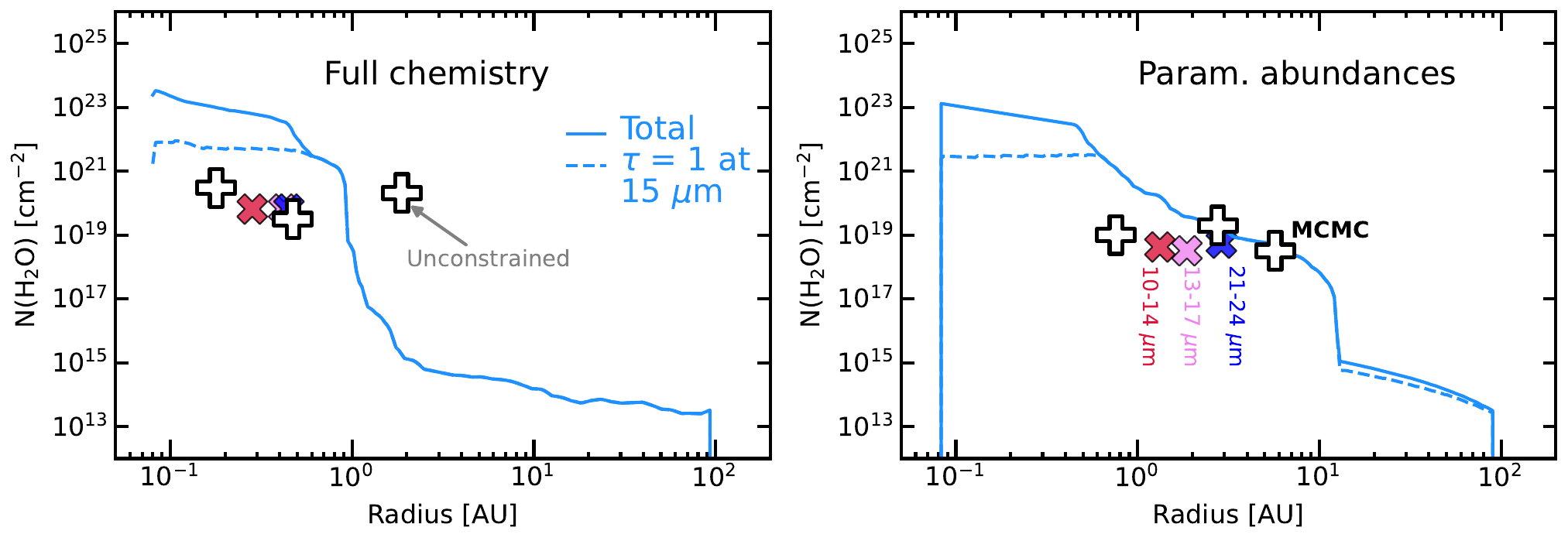}
    \caption{Vertically integrated \ce{H2O} column density as a function of radius for the models with $f_\ell=0.999$ with the full chemistry (left panel) and with parameterized abundances (right panel). The solid lines show the total model column density and the dashed lines show the model column density integrated up to the dust $\tau=1$ surface at 15 $\mu$m. The red, pink, and blue crosses represent the retrieved $N$ and $R_{\rm eq}$ from the single-temperature slab fits in the 10-14, 13.5-17.5, and 21-24 $\mu$m region. The white plus symbols represent the retrieved $N$ and $R_{\rm eq}$ values from the 3-temperature-component MCMC routine.}
    \label{fig:N_rad_fl999}
\end{figure*}

\begin{figure*}
    \centering
    \includegraphics[width=0.95\linewidth]{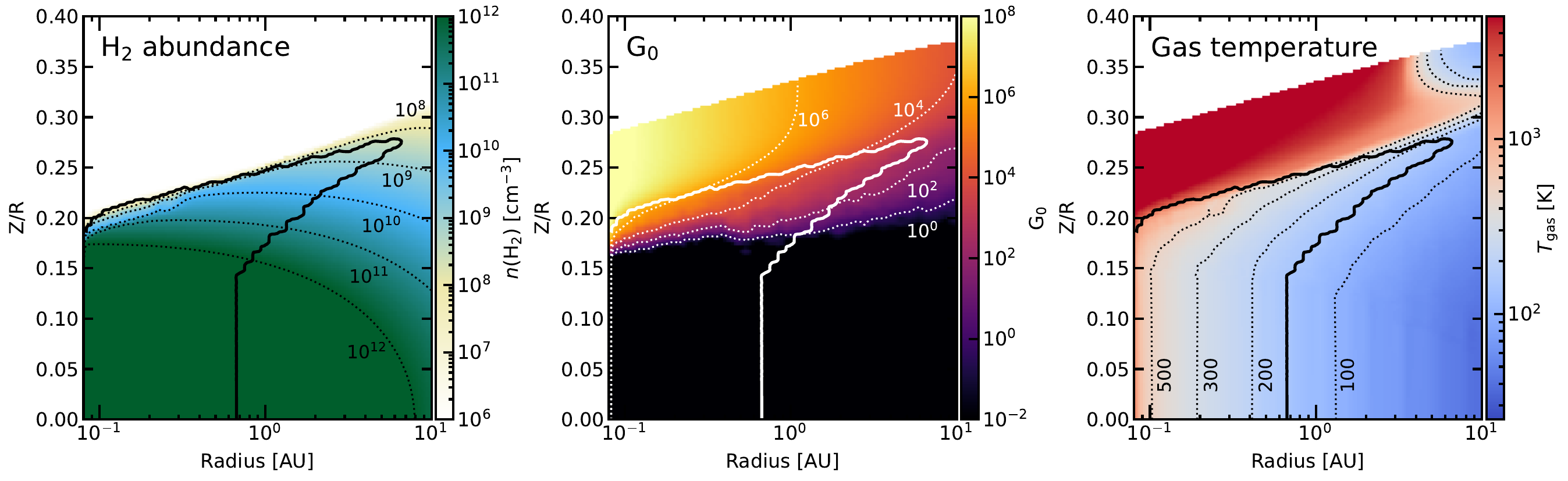}
    \caption{Maps of the \ce{H2} density, UV radiation field $G_0$, and gas temperature of the fiducial chemistry model. The solid black (and white in the second panel) contours represent the \ce{H2O} abundance structure of the fiducial parameterized model (see Fig. \ref{fig:abu_fl9}).}
    \label{fig:H2_G0_Tgas}
\end{figure*}

\begin{figure*}
    \centering
    \includegraphics[width=0.95\linewidth]{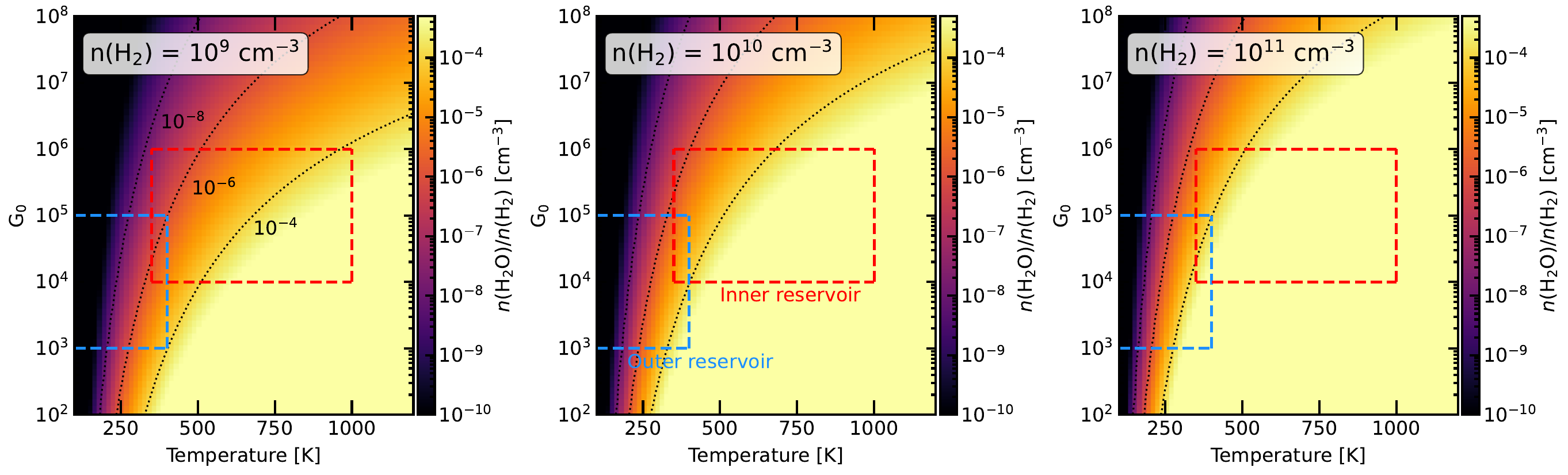}
    \caption{\ce{H2O} abundance as a function of gas temperature and UV radiation field, following a simple chemical model (see Sect. \ref{app:tgas}). The red and blue boxes represent the conditions within the inner and outer \ce{H2O} reservoir.}
    \label{fig:nH2O}
\end{figure*}

\begin{figure*}
    \centering
    \includegraphics[width=0.95\linewidth]{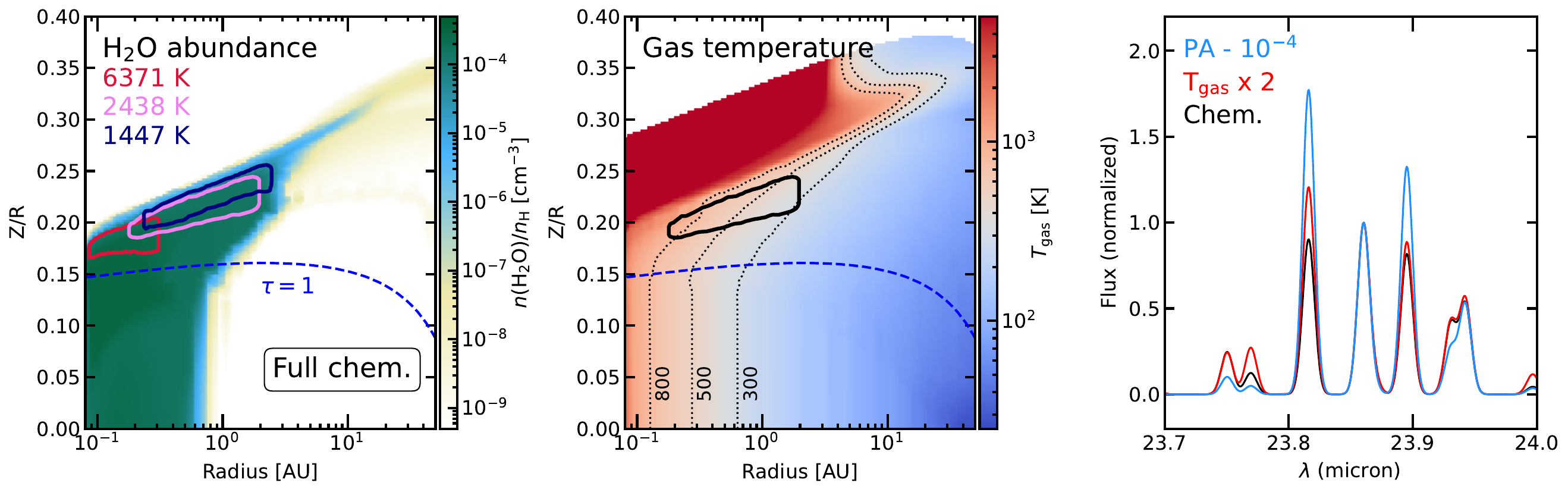}
    \caption{\ce{H2O} abundance, gas temperature, and \ce{H2O} spectrum between 23.7 and 24 $\mu$m for a model with a gas temperature increased by a factor 2 with respect to the fiducial chemistry model. In the first panel, red, pink, and blue contours represent the 70\% emitting regions of the \ce{H2O} 17$_{7,10}$ -- 16$_{4,13}$ ($E_{\rm up} = 6371 K$), 11$_{3,9}$ -- 10$_{0,10}$ ($E_{\rm up} = 2438 K$; also shown in black in the second panel), and 8$_{3,6}$ -- 7$_{0,7}$ ($E_{\rm up} = 1447 K$) lines, respectively. The dust $\tau=1$ surface at 15 $\mu$m is indicated with a blue dashed line in the first and second panel. In the third panel, the fiducial chemistry (chem.) model is shown in black, the model with increased temperature is shown in red, and the fiducial parameterized model is shown in blue.}
    \label{fig:abu_Tgas}
\end{figure*}

\end{appendix}

\end{document}